\documentclass{jfm}

\usepackage{cite}
\usepackage{url}

\usepackage{graphicx}
\usepackage{natbib}
\usepackage{amssymb,amsmath}

\usepackage{graphicx, amsmath, esint, bm}
\usepackage{epstopdf, epsfig}

\newcommand{\I}{\mathrm{i}}
\newcommand\D{\mbox{d}}

\newcommand{\p}{\partial}

\newcommand{\C}{\mathbb C}
\newcommand{\e}{\eqref}

\newcommand \bfr{{\bf r}}

\ifCUPmtlplainloaded \else
  \checkfont{eurm10}
  \iffontfound
    \IfFileExists{upmath.sty}
      {\typeout{Found AMS Euler Roman fonts on the system,
                   using the 'upmath' package.}%
       \usepackage{upmath}}
      {\typeout{Found AMS Euler Roman fonts on the system, but you
                   dont seem to have the}%
       \typeout{'upmath' package installed. JFM.cls can take advantage
                 of these fonts, if you use 'upmath' package.}%
       \providecommand\upi{\upi}%
      }
  \else
    \providecommand\upi{\upi}%
  \fi
\fi

\ifCUPmtlplainloaded \else
  \checkfont{msam10}
  \iffontfound
    \IfFileExists{amssymb.sty}
      {\typeout{Found AMS Symbol fonts on the system, using the
                'amssymb' package.}%
       \usepackage{amssymb}%
       \let\le=\leqslant  
         
      }{}
  \fi
\fi

\ifCUPmtlplainloaded \else
  \IfFileExists{amsbsy.sty}
    {\typeout{Found the 'amsbsy' package on the system, using it.}%
     \usepackage{amsbsy}}
    {\providecommand\boldsymbol[1]{\mbox{\boldmath $##1$}}}
\fi

\shorttitle{Non-Canonical Hamiltonian Structure and Poisson Bracket%
} \shortauthor{A. I. Dyachenko, P. M. Lushnikov, and V. E. Zakharov}

\title[Non-Canonical Hamiltonian Structure and Poisson Bracket]{Non-Canonical Hamiltonian Structure and Poisson Bracket for $2$D Hydrodynamics with Free Surface}

\author[A. I. Dyachenko,   P. M. Lushnikov, and V. E.
Zakharov]{ A. I. Dyachenko$^{1}$,
           P. M. Lushnikov$^{1,2}
           $\thanks{Email address for correspondence: plushnik@math.unm.edu} and
           V. E. Zakharov$^{1,3}$
 }

\affiliation{ \aff{1}Landau Institute for Theoretical Physics,
Russia  \aff{2}Department
of Mathematics and Statistics, University of New Mexico,
Albuquerque, NM 87131, USA \aff{3}Department of Mathematics,
University of Arizona, Tucson, AZ 85721, USA }
\affiliation{ $^{1}$Landau Institute for Theoretical Physics, Russia\\[\affilskip]
 $^{2}$University of New
Mexico, Albuquerque, NM 87131, USA \\[\affilskip]
$^{3}$Department of Mathematics, University of Arizona, Tucson, AZ
85721, USA }
\date{%Printed
March 29, 2019}

\begin{document}

\maketitle

\begin{abstract}
We consider Euler equations for potential flow of ideal
incompressible fluid with a free surface and infinite depth in two
dimensional geometry.   Both gravity forces and surface tension are
taken int account. A time-dependent conformal mapping is used which
maps a lower complex half plane of  the auxiliary complex variable
$w$    into a fluid's area  with the real line of $w$ mapped into
the free fluid's surface. We reformulate  the exact Eulerian
dynamics through a non-canonical nonlocal Hamiltonian structure for
a pair of the Hamiltonian variables. These two variables are the
imaginary part of the conformal map and the fluid's velocity
potential both evaluated of fluid's free surface.    The
corresponding Poisson bracket is non-degenerate,  i.e. it does not
have any Casimir invariant. Any two functionals of the conformal
mapping   commute with respect to the Poisson bracket.       New
Hamiltonian structure is a generalization of the canonical
Hamiltonian structure of Ref. V.E. Zakharov, J. Appl. Mech. Tech.
Phys. 9, 190 (1968) which is valid only for solutions for which the
natural surface parametrization is single valued, i.e. each value of
the horizontal coordinate corresponds only to a single point on the
free surface. In contrast, new non-canonical Hamiltonian equations
are valid for arbitrary nonlinear solutions (including
multiple-valued natural surface parametrization) and are equivalent
to Euler equations.  We  also consider a generalized hydrodynamics
with the additional physical terms in the Hamiltonian beyond the
Euler equations.  In that case we identified  powerful reductions
which allowed to find  general classes of particular solutions.
\end{abstract}

\section{Introduction and basic equations}
\begin{figure}
\includegraphics[width=0.9859\textwidth]{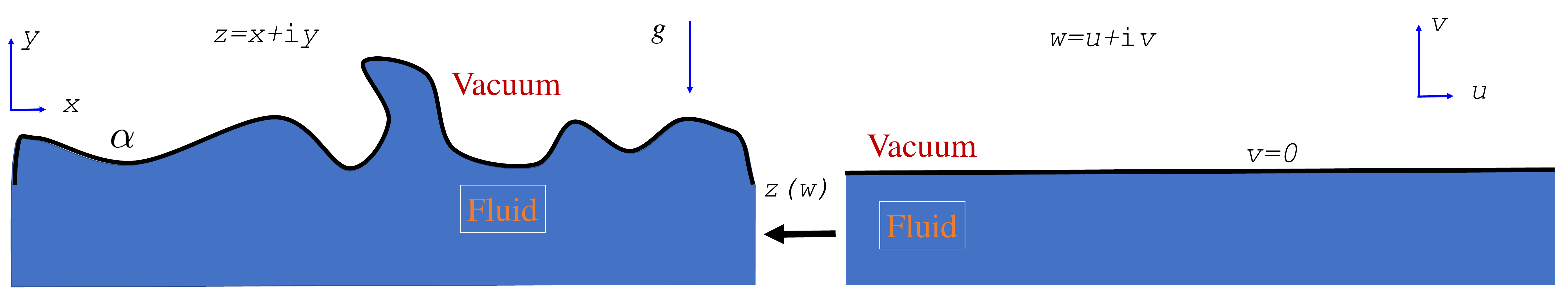}%{figconformal.pdf}
\caption{ Dark area represents the domain occupied by fluid in the
physical plane $z=x+\I y $ (left) and the same domain in   $w=u+\I
v$ plane  (right). Thick solid lines correspond to the fluid's free
surface in both planes.} \label{fig:schematic1}
\end{figure}

We study two-dimensional potential motion of ideal incompressible fluid with free surface of infinite depth. Fluid occupies the infinite region $-\infty < x < \infty$ in the horizontal direction $x$ and extends down to $y\to -\infty$ in the vertical direction $y$   as schematically shown on the left panel of Fig. \ref{fig:schematic1}. The time-dependent fluid free surface is represented in the parametric form as
\begin{equation} \label{xyu}
x=x(u,t), \ y=y(u,t)
\end{equation}
with the parameter  $u$  spanning the range $-\infty<u<\infty$ such that %
\begin{equation} \label{xyinfinity}
x(u,t)\to \pm \infty \ \text{and} \ y(u,t)\to 0 \ \text{as}  \ u\to \pm \infty.
\end{equation}
 We assume that the free surface does not have self-intersection, i.e. $\bfr(u_1,t)\neq\bfr(u_2,t)$ for any $u_1\neq u_2$. In other words, the free surface is the simple plane curve. Here $\bfr(u,t)\equiv(x(u,t),y(u,t))$.

In  the particular case when the free surface can be represented by a single-valued function of $x$,%
\begin{equation} \label{etadef}
 y = \eta(x,t) ,  \end{equation}
one can also represent domain occupied by the fluid as $-\infty <y \le \eta$ and  $-\infty < x < \infty.$     Such single-valued case has been widely considered (see e.g. Ref. \cite{StokerBook1957}). We however do not restrict to that particular case which is recovered by choosing $u=x$.

Potential  motion implies that a velocity ${\bf v}$    of fluid is determined by a
velocity potential $\Phi(\bfr,t)$ as ${\bf v}= \nabla \Phi$ with $\nabla\equiv(\frac{\p}{\p x},\frac{\p}{\p y})$.  The
incompressibility condition $\nabla \cdot {\bf v} = 0$ results in
the
Laplace equation
\begin{align} \label{laplace}
\nabla^2 \Phi = 0
\end{align}
inside fluid. Eq. \e{laplace} is supplemented with  decaying boundary condition (BC) at infinity in the horizontal direction,  %
\begin{equation} \label{phiinfinity}
\nabla\Phi\to 0 \ \text{for } \ |x|\to \infty,
\end{equation}
and a vanishing of the normal velocity  the  fluid's bottom,

\begin{equation} \label{vbottom}
\left .\frac{\p \Phi}{\p n}\right |_{y\to-\infty}=0.
\end{equation}
Without loss of generality BCs \e{phiinfinity} and \e{vbottom} can be replaced by Dirichlet BC

\begin{equation} \label{Dirichlet1}
\Phi\to 0 \  \text{at} \ |\bfr|\to \infty.
\end{equation}

BCs at the free surface  are time-dependent and consist of kinematic and dynamic BCs.
Kinematic BC ensures that  free surface moves with the normal velocity component $v_n$ of fluid particles at the free surface.   Motion of the free surface is determined by  time derivatives of the parameterization  \e{xyu}  and kinematic BC is given by a projection into normal directions as
\begin{equation} \label{kinematicu0}
{\bf n}\cdot\left(x_t,y_t \right )=v_n\equiv{\bf n}\cdot\nabla \Phi|_{x=x(u,t),\ y=y(u,t)},
\end{equation}
where %
\begin{equation} \label{ndef}
{\bf n}=\frac{(-y_u,x_u)}{(x_u^2+y_u^2)^{1/2}}
\end{equation}
is the outward unit normal vector to the free surface and subscripts here and below means partial derivatives, $x_t\equiv\frac{\p x(u,t)}{\p t}$ etc.

Eqs. \e{kinematicu0} and \e{ndef} result in a compact expression%
\begin{equation} \label{kinematicu}
y_tx_u  -x_t  y_u =[x_u\Phi_y-y_u\Phi_x]|_{x=x(u,t),\ y=y(u,t)}
\end{equation}
for the kinematic BC.

 Tangential component of the vector $\bfr_t=\left(x_t,y_t \right)$ is not  fixed by kinematic BC \e{kinematicu} but can be chosen at our convenience. E.g., one can define $u$ to be the Lagrangian coordinate of fluid particles at the free surface (fluid particles once on the free surface never leave it). Then tangential component of $\bfr_t$ would coincide with  the tangential component of  $\nabla \Phi|_{x=x(u,t),\ y=y(u,t)}$. Another possible choice is to choose $u$ to be the  arclength along the free surface. However, we use neither Lagrangian or arclength formulation below. Instead, throughout the paper we use the conformal variables for the free surface parameterization as described below in Section \ref{sec:conformalmapping}. Another particular form of   \e{xyu}  is given by Eq. \e{etadef}, which corresponds to choosing $u=x$ (as mentioned above, it is possible  only if $\eta(x,t)$ is the single-valued function of $x$). In that case Eq. \e{ndef} is reduced to ${\bf n}=(-\eta_x,1)(1+\eta_x^2)^{-1/2}$ and kinematic BC Eq. \e{kinematicu} is given by%
\begin{equation} \label{kinematiceta}
\eta_t=(1+\eta_x^2)^{1/2}v_n=\left .\left (-\eta_x\Phi_x+\Phi_y\right )\right|_{\ y=\eta(x,t)}.
\end{equation}
This form of kinematic BC has been widely used (see e.g. Ref. \cite{StokerBook1957}).

 A dynamic BC, which is the time-dependent Bernoulli equation (see e.g. \citet{LandauLifshitzHydrodynamics1989})
at the free surface, is given by%
\begin{align} \label{dynamic1}
\left.\left(\Phi_t +
 \dfrac{1}{2}\left(\nabla \Phi\right)^2+gy\right)\right|_{x=x(u,t),\ y=y(u,t)}  = -P_\alpha,
\end{align}
 where $g$ is the acceleration due to gravity and%
\begin{equation} \label{Palpha}
P_\alpha=-\frac{\alpha(x_uy_{uu}-x_{uu}y_u) }{(x_u^2+y_u)^{3/2}}
\end{equation}
is the pressure jump  at the free
surface due to the surface tension coefficient $\alpha$. Here without loss of generality we assumed that pressure is zero above the free surface (i.e. in vacuum).
All results below apply both to the surface gravity wave case ($g>0$) and the Rayleigh-Taylor problem $(g<0)$. Below we also consider a particular case $g=0$ when inertia forces  well exceed
 gravity force. For the  case of single-valued parameterization \e{etadef}, Eq. \e{Palpha} is reduced to the well-known expression (see e.g. Ref. \cite{Zakharov1968})%
\begin{equation} \label{Palphaeta}
P_\alpha=-\alpha\frac{\p }{\p
x}[\eta_{x}(1+\eta_x^2)^{-1/2}]=-\alpha\eta_{xx}(1+\eta_x^2)^{-1/2}.
\end{equation}
Eqs.  \e{dynamic1} and \e{Palpha}, together with decaying BCs \e{xyinfinity} and \e{phiinfinity}, imply that a Bernoulli constant (generally located at right hand side (r.h.s) of Eq. \e{dynamic1}) is zero.

Eqs. \e{xyu},\e{xyinfinity},\e{laplace}-\e{ndef},\e{dynamic1} and \e{Palpha} form a closed set of equations which is equivalent to Euler equations for dynamics of ideal fluid with  free surface for any chosen free surface parameterization \e{xyu}. Here at each moment of time $t$, Laplace Eq.  \e{laplace} has to be solved with Dirichlet BC   %
\begin{equation} \label{dirichletpsi}
\psi(x,t)\equiv\Phi(\bfr,t)|_{x=x(u,t),\ y=y(u,t)}
\end{equation}
and  BCs \e{phiinfinity}, \e{vbottom}. That boundary value problem has the unique solution. The knowledge of $\Phi(\bfr,t)$ allows to find the normal velocity $v_n$ at the free surface as in Eq. \e{kinematicu}. It can be interpreted as finding the Dirichlet-Neumann operator for the Laplace Eq. \e{laplace} \cite{CraigSulemJCompPhys1993}. Then one can advance in time to find new value of $\psi(x,t)$ from Eqs.  \e{kinematicu} and \e{dynamic1} using that %
\begin{equation} \label{psiteq}
\psi_t=\left [\Phi_t+x_t\Phi_x+y_t \Phi_y\right ]|_{x=x(u,t),\ y=y(u,t)}
\end{equation} as well as evolve a parameterization  \e{xyu} and so on. Here Eq. \e{psiteq} results from the definition \e{dirichletpsi}.

The set \e{xyu},\e{xyinfinity},\e{laplace},\e{Dirichlet1},\e{ndef} and \e{dynamic1} preserves the total energy%
\begin{equation} \label{Hamilttotal}
H=K+P,
\end{equation}
where %
\begin{equation} \label{KineticEnergy}
K=\frac{1}{2}\int \limits_\Omega{(\nabla \Phi)^2}\D x \D y
\end{equation}
is the kinetic energy
and
\begin{equation} \label{PotentialEnergy}
P=g\int \limits_\Omega y\,\D x \D y-g\int \limits_{y\le 0} y\,\D x \D y+\alpha\int\limits^{\infty}_{-\infty}\left (\sqrt{x_u^2+y_u^2}-x_u\right )\D u
\end{equation}
is the potential energy. Here $\D x \D y$ is the element of fluid volume (more precisely it is the fluid's area because we restricted to 2D fluid motion with the third spatial  dimension being trivial), $\Omega$ is the area occupied by the fluid which extends down to $y\to-\infty$ in the vertical direction. The term $g\int \limits_{y\le 0} y\,\D x \D y$ corresponds to the gravitational energy of unperturbed fluid (flat free surface) and it is subtracted from the integral over $\Omega$  to ensure that the total contribution of the gravitational energy, $g\int _\Omega y\,\D x \D y-g\int \limits_{y\le 0} y\,\D x \D y$, is finite. In other words, one can understand these two terms    as the limit $h\to\infty$ and $L\to \infty,$ where $h$ is the fluid depth with the bottom at  $y=-h$ and $L$ is the horizontal extend of the fluid. Then $g\int \limits_{y\le 0} y\,\D x \D y=-\frac{g h^2L}{2} ,$ where using this expression below we assume taking the limits $h\to\infty$ and $L\to \infty.$  The surface tension energy $\alpha\int^{\infty}_{-\infty}\left (\sqrt{x_u^2+y_u^2}-x_u\right )\D u$ in Eq.  \e{PotentialEnergy} is determined by the arclength of free surface with $-x_u$ term added to ensure that the surface energy is zero for unperturbed fluid with $y\equiv 0$.

If we introduce the vector field ${\bf F}=\hat y y^2/2$ with $\hat
y$ being the unit vector in positive $y$ direction, then the
gravitational energy in Eq.  \e{PotentialEnergy} takes the following
form $g\int _\Omega \nabla \cdot {\bf F}\,\D x \D y- \frac{g
h^2L}{2}|_{h,L\to \infty}$. By the  divergence theorem of vector
analysis
(in our 2D case it can be also reduced to the Green's theorem) this gravitational energy is converted into
 the surface integral $g\int _{\p \Omega}  {\bf F}\cdot {\bf n}\,\D s+\frac{g h^2L}{2}|_{h,L\to \infty}$
  (line integral in 2D over arclength $\D s=\sqrt{x_u^2+y_u^2}\D u$ with $\p \Omega$ being the boundary of $\Omega$) which together with Eq.  \e{ndef}  results in  %
\begin{equation} \label{PotentialEnergy2}
P=\frac{g}{2}\int \limits_{-\infty}^{\infty} y^2\,x_u\D  u+\alpha\int\limits^{\infty}_{-\infty}\left (\sqrt{x_u^2+y_u^2}-x_u\right )\D u.
\end{equation}

In the simplest case of the single-valued surface parametrization Eq. \e{etadef}, Eqs. \e{KineticEnergy} and \e{PotentialEnergy2} take the simpler forms    %
\begin{equation} \label{KineticEnergy3}
K=\frac{1}{2}\int \limits_{-\infty}^\infty\D x\int\limits^\eta_{-\infty}{(\nabla \Phi)^2} \D y
\end{equation}
and
\begin{equation} \label{PotentialEnergy3}
P=\frac{g}{2}\int \limits_{-\infty}^{\infty} \eta^2\D x+\alpha\int\limits^{\infty}_{-\infty}\left (\sqrt{1+\eta_x^2}-1\right )dx,
\end{equation}
respectively.

It was proved in Ref. \cite{Zakharov1968} that $\psi$ and $\eta$ for the single-valued surface parametrization   \e{etadef} satisfy the canonical Hamiltonian system%
\begin{equation} \label{Hamiltpsieta}
\frac{\p \eta}{\p t}=\frac{\delta H}{\delta \psi}, \quad \frac{\p
\psi}{\p t}=-\frac{\delta H}{\delta \eta}
\end{equation}
with $H$ given by Eqs. \e{Hamilttotal}, \e{KineticEnergy3} and
\e{PotentialEnergy3}. The Hamiltonian formalism of Ref.
\cite{Zakharov1968} has been widely used for water waves, see e.g.
Refs. \cite{ZLF1992,KharifPelinovskyEurJMechB2003} for review as
well as it was generalized to the dynamics of the interface between
two fluids \cite{KuznetsovLushnikov1995}. In this paper we show that
for the general ``multivalued'' case of the parametrization \e{xyu},
the system of dynamical Eqs. \e{kinematicu} and \e{dynamic1} for
$x(u,t),$ $y(u,t)$ and $\psi(u,t)$ also has a Hamiltonian structure
if we additionally assume that $x(u,t)$ and $y(u,t)$ are defined
from the conformal map of Section \ref{sec:conformalmapping}.
However, that structure is non-canonical with the non-canonical
Poisson bracket and depends on the choice of the parametrization of
the surface.

Apparently, the system
  \e{kinematicu} has infinite number of degrees of freedom. The most important feature of integrable systems
   is the existence of ``additional"
constants of motion which are different from ``natural" motion constants (integrals) (see Refs. \cite{GardnerGreeneKruskalMiuraPRL1967,ZakharovShabatJETP1972,ArnoldClassicalMechanics1989,ZakharovFaddeevFunktAnalPril1971eng,NovikovManakovPitaevskiiZakharovbook1984}). For system  \e{Hamiltpsieta} the natural integrals are the energy $H$  \e{Hamilttotal}, the total mass of fluid, %
\begin{equation} \label{Mass}
M=\int\limits^\infty_{-\infty}\eta(x,t)\D x,
\end{equation}
and the horizontal component of the momentum, %
\begin{equation} \label{Pxdef}
P_x=\int\limits^\infty_{-\infty}\D x\int\limits^\eta_{-\infty}\frac{\p \Phi}{\p x}\D y.
\end{equation}

$\Phi$ is the harmonic function inside fluid because it satisfies the Laplace Eq. \e{laplace}. The harmonic conjugate of $\Phi$ is a  stream  function $\Theta$ defined by%
\begin{equation} \label{Thetadef}
\Theta_x=-\Phi_y \ \text{and} \ \Theta_y=\Phi_x.
\end{equation}
Similar to Eq. \e{Dirichlet1}, we set without loss of generality zero  Dirichlet BC for $\Theta$ as

\begin{equation} \label{Dirichlet2}
\Theta\to 0 \  \text{at} \ |\bfr|\to \infty.
\end{equation}

 We define a complex velocity potential $\Pi(z,t)$~as%
\begin{equation} \label{ComplexPotentialdef}
\Pi=\Phi+\I \Theta,
\end{equation}
 where %
\begin{equation} \label{zdef}
z=x+\I y
\end{equation}
is the complex coordinate. Then Eqs. \e{Thetadef} turn into
Cauchy-Riemann equations ensuring the analyticity of $\Pi(z,t)$ in the domain of $z$ plane occupied by the fluid (with the free fluid's boundary defined by Eqs. \e{xyu} and \e{xyinfinity}).
A physical velocity with the components
$v_x$ and $v_y$ (in $x$ and $y$ directions, respectively) is recovered from $\Pi$ as $\frac{d\Pi}{dz}=v_x-\I v_y$.

Using
$\Theta_y=\Phi_x$ from Eq. \e{Thetadef}, we immediately convert the horizontal momentum \e{Pxdef}
into $P_x=\int^\infty_{-\infty}\Theta\, \D x$ through integration by parts and Eq. \e{Dirichlet2} which results in
\begin{equation} \label{Pxdef2}
P_x=\int\limits^\infty_{-\infty}\Theta(x(u,t),y(u,t),t)x_u(u,t)\D u.
\end{equation}
Eq. \e{Pxdef2} is also valid for the general multi-valued case
(contrary to Eq. \e{Pxdef} which requires the particular
parametrization \e{etadef}). To check that we replace Eq. \e{Pxdef}
by $P_x=\int_\Omega\Theta_y\D  x\D y=\int_\Omega\nabla \cdot{\bf
F}\,\D x\D y$ with ${\bf F}=\hat y \Theta$ and, similar to the
derivation of Eq. \e{PotentialEnergy2}, we then obtain Eq.
\e{Pxdef2} from the divergence theorem and Eq. \e{ndef}.

One can use Eqs. \e{Thetadef} to obtain the equivalent form of $P_x$
as  $P_x=\int_\Omega\Phi_x\D  x\D y=\int_\Omega\nabla \cdot{\bf
F}\,\D x\D y$ with ${\bf F}=\hat x \Phi. $ Then  the divergence
theorem together with Eq.  \e{ndef} results  in
\begin{equation} \label{Pxdef3}
P_x=-\int\limits^\infty_{-\infty}\psi(x(u,t),t)y_u(u,t)\D u.
\end{equation}

In a similar way, a vertical component of momentum is given by
\begin{equation} \label{Pydef2a}
P_y=\int\limits^\infty_{-\infty}\D x\int\limits^\eta_{-\infty}\frac{\p \Phi}{\p y}\D y=\int\limits^\infty_{-\infty}\psi\, \D x,
\end{equation}
where we used integration by part and Eqs. \e{Dirichlet1}  and \e{dirichletpsi}.  $P_y$ is the integral of motion only for the zero gravity case, $g=0.$
 A change of integration variable in Eq. \e{Pydef2a} results in
\begin{equation} \label{Pydef2}
P_y=\int\limits^\infty_{-\infty}\psi\, x_u\D u.\end{equation} Eq.
\e{Pydef2} is also valid for the general multi-valued case. To check
that we define in the general case that $P_y=\int_\Omega\Phi_y\,\D
x\D y=\int_\Omega\nabla \cdot{\bf F\,}\D x\D y$ with ${\bf F}=\hat y
\Phi$ and, similar to the derivation of Eq. \e{PotentialEnergy2}, we
obtain Eq.  \e{Pydef2}  from the divergence theorem and Eq.
\e{ndef}.

One can use Eqs. \e{Thetadef} to obtain the equivalent form of $P_y$
as  $P_y=-\int_\Omega\Theta_x\D  x\D y=\int_\Omega\nabla \cdot{\bf
F}\,\D x\D y$ with ${\bf F}=-\hat x\Theta. $ Then  the divergence theorem together with Eq.  \e{ndef} results in
\begin{equation} \label{Pydef3}
P_x=\int\limits^\infty_{-\infty}\Theta(x(u,t),y(u,t),t)y_u(u,t)\D u.
\end{equation}

For the parametrization \e{xyu}, Eq. \e{Mass}  is replaced by
$M=\int_\Omega\D  x\D y-\int_{y\le0}\D  x\D y=\int_\Omega\nabla
\cdot{\bf F}\D x\D y-hL|_{h,L\to \infty}$ with ${\bf F}=\hat y y$.
Similar to derivation of Eq. \e{PotentialEnergy2} we then use the
divergence theorem and  Eq.  \e{ndef} to obtain that
\begin{equation} \label{Mass2}
M=\int\limits^\infty_{-\infty}y(u,t) x_u(u,t)\D u.
\end{equation}
%

%In this paper we show that both the system
%  \e{kinematicu} and more general multi-valued case have  ???  a number of additional constants of motion. We cannot so far determine/estimate a total number of these constants. Instead we show examples of initial data such that the system has almost obvious, very simply constructed additional constants. We must stress that that the number of known additional constants depends so far on the choice of initial data and can be made arbitrary large for the specific choices of initial data.

%\newpage

In this paper we develop a Hamiltonian formalizm for the general
multi-valued case compare with single-valued case established in
Ref. \cite{Zakharov1968}. Plan of the paper is the following.
 In Section \ref{sec:conformalmapping} we introduce the conformal variables as the particular case of
 the general parametrization  \e{xyu}. In Section \ref{sec:Hamiltonian formalizm} we introduce the Hamiltonian formalism for system
 \e{xyu},\e{xyinfinity},\e{laplace}-\e{ndef} and \e{dynamic1} with the nonlocal non-canonical symplectic form and the corresponding Poisson bracket.
 Section \ref{Dynamicalequations} provides the explicit expression for the Hamiltonian equations resolved with respect to time derivatives.  Section \ref{DynamicalequationsComplex} rewrites these dynamic equations in the complex form and introduce another complex unknowns $R$ and $V$.
Section \ref{ref:Generalizedhydrodynamicsintegrability} introduce a
generalization of the Hamiltonian of Euler equations with free
surface to include additional physical effects such as the
interaction of dielectric fluids with electric field and two fluid hydrodynamics of superfluid Helium with a free surface. It is shown
that these equations allows very powerful reductions which suggests
a complete integrability. Section \ref{sec:conclusion} provides a
summary of obtained results and discussion on future directions.

\section{Conformal mapping}
\label{sec:conformalmapping}

To choose a convenient version of the general parametrization  \e{xyu}, we consider  the time-dependent conformal mapping
\begin{equation} \label{zwdef}
z(w,t)=x(u,v,t)+\I y(u,v,t)
\end{equation}
of the lower complex half-plane $\mathbb{C}^-$ of the auxiliary complex variable
\begin{equation} \label{wdef}
w\equiv u+\I v, \quad -\infty<u<\infty,
\end{equation}
into the area in $(x,y)$ plane occupied by the fluid. Here the real line $v=0$ is mapped into the fluid free surface (see Fig. \ref{fig:schematic1}) and $\mathbb{C}^-$ is defined by the condition  $-\infty<v<0$.  The function $z(w,t)$ is the analytic function of   $w\in\mathbb{C^-}$. The conformal mapping \e{zwdef} at $v=0$ provides a particular form of the free surface parameterization  \e{xyu} for the parameter $u. $

 The conformal mapping \e{zwdef} ensures that the function $\Pi(z,t)$~ %
\e{ComplexPotentialdef} transforms into $\Pi(w,t)$ which is analytic function of $w$ for $w\in\mathbb{C^-}$ (in the bulk of
fluid).
Here and below we abuse the notation and use the  same symbols for functions of either  $w$ or $z$     (in other words, we  assume that e.g. $\tilde \Pi(w,t)= \Pi(z(w,t),t) $ and remove $\tilde ~$ sign).
The conformal transformation    \e{zwdef} also ensures   Cauchy-Riemann equations $
\Theta_u=-\Phi_v, \quad \Theta_v=\Phi_u $
  in $w$ plane.

The idea of using time-dependent conformal transformation like
\e{zwdef} to address systems equivalent/similar to  Eqs.
\e{xyu},\e{xyinfinity},\e{laplace}-\e{ndef} and \e{dynamic1}   was
exploited by several authors
including~\cite{Ovsyannikov1973,MeisonOrzagIzraelyJCompPhys1981,TanveerProcRoySoc1991,TanveerProcRoySoc1993,DKSZ1996,ChalikovSheininAdvFluidMech1998,ChalikovSheininJCompPhys2005,ChalikovBook2016,ZakharovDyachenkoVasilievEuropJMechB2002}.
We follow~\cite{DKSZ1996} to recast the system
\e{xyu},\e{xyinfinity},\e{laplace}-\e{ndef} and \e{dynamic1} into
the equivalent form for $x(u,t), \ y(u,t)$ and $\psi(u,t)$ at the
real line $w=u$ of the complex plane $w$ using the conformal
transformation  \e{zwdef}.
We show that the kinematical BC takes the form  %
\begin{equation}\label{fullconformal1}
 y_tx_u  -x_t  y_u =- \hat {\mathcal H} \psi_u,
\end{equation}
where
\begin{equation} \label{Hilbertdef}
\hat {\mathcal H} f(u)=\frac{1}{\upi} \text{p.v.}
\int^{+\infty}_{-\infty}\frac{f(u')}{u'-u}\D u'
\end{equation}
is the Hilbert transform (\cite{Hilbert1905})  with $\text{p.v.}$
denoting a Cauchy principal value of integral. The dynamic BC takes
the form \begin{equation}\label{fullconformal2} \psi_t y_u - \psi_u
y_t + gyy_u =- \hat {\mathcal H} \left (\psi_t x_u - \psi_ux_t +
gyx_u \right )-\alpha\frac{\p  }{\p u}\frac{x_u}{|z_u|}+\alpha\hat
{\mathcal H} \frac{\p  }{\p u}\frac{y_u}{|z_u|},
\end{equation}
where $x(u,t)$ is expressed through $y(u,t) $ as follows
\begin{equation} \label{xHy}
\tilde    x\equiv x-u=-\hat {\mathcal H}y
\end{equation}
(see Eq. \e{ReImHilbert} of Appendix \ref{sec:AppA}  for the
justification of Eq.  \e{xHy} as well as the complimentary
expression  $\hat {\mathcal H} \tilde x=y$).

Eq. \e{xHy} exemplifies the general relation between   the harmonically conjugated functions in $\C^-$ as was first obtained by David Hilbert  (\cite{Hilbert1905}). The particular case of  Eq. \e{xHy} results from the analyticity of   $z(w,t)$   for $w\in\mathbb{C}^-$ which implies that $\tilde x$ and $y$ are harmonically conjugated functions for $w\in\mathbb{C}^-.$   Similarly, $\Pi(w,t)$ \e{ComplexPotentialdef} is also analytic function for $w\in\mathbb{C}^-$ which results in %
\begin{equation} \label{PhiTheta1}
\Theta |_{w=u}= \hat {\mathcal H} \psi, \quad \psi =- \hat {\mathcal
H} \Theta|_{w=u} \ \text{for} \ w=u.
\end{equation}

We notice that left hand side (l.h.s.) of Eq. \e{fullconformal1}
is the same as l.h.s of Eq. \e{kinematicu}
multiplied by $(x_u^2+y_u^2)^{1/2}=|z_u|$. R.h.s. of Eq. \e{kinematicu}
multiplied by $(x_u^2+y_u^2)^{1/2}$ is given by $\Phi_v|_{v=0}= -\Theta_u|_{v=0}$  (which is the normal velocity $v_n$ to the surface in $w$ plane multiplied by the Jacobian    $x_u^2+y_u^2$ of the conformal transformation \e{zwdef}, see e.g. Refs. \cite{DKSZ1996,DyachenkoLushnikovKorotkevichPartIStudApplMath2016}). Then using Eqs. \e{dirichletpsi} and \e{PhiTheta1}, we obtain Eq. \e{fullconformal1}.

Eq. \e{fullconformal2} can be also obtained from Eqs. \e{laplace}-\e{ndef},\e{dynamic1},\e{Palpha} and \e{dirichletpsi} by the change of variables \e{zwdef}.
We do not provide it here to avoid somewhat bulky  calculations. Instead, we  derive  both Eqs. \e{fullconformal1} and \e{fullconformal2} from Hamiltonian formalism in Section \ref{sec:Hamiltonian formalizm}.
See also Appendix A.2 of Ref. \cite{DyachenkoLushnikovKorotkevichPartIStudApplMath2016} for detailed derivation of similar Eqs.
for a case of the periodic BCs along $x$ instead of decaying BCs  \e{xyinfinity} and \e{phiinfinity}.

%Notice that Eqs. \e{fullconformal1} and \e{fullconformal2} are valid for any simply-connected domain in ($(x,y)$ plane??? (check that, $\hat {\mathcal H}$ generally would have to be replaced by something else????).

We now transform  the kinetic energy  \e{KineticEnergy} into the integral over the real line $w=u$. The Laplace Eq. \e{laplace} implies that we can apply the Green's formula to Eq. \e{KineticEnergy}  as   $K=\frac{1}{2}\int _\Omega{\nabla\cdot(\Phi\nabla \Phi)}\D x \D y=\frac{1}{2}\int _{\p\Omega}{\psi v_n}\D s=\frac{1}{2}\int _{\p\Omega}{\psi v_n} \sqrt{x_u^2+y_u^2}\D u$. Using Eqs.  \e{ndef}, \e{dirichletpsi}, \e{PhiTheta1}  and rewriting $v_n$ in conformal variable $w$
 (see e.g. Appendix A.1 of Ref. \cite{DyachenkoLushnikovKorotkevichPartIStudApplMath2016}  for the explicit expressions on the respective derivatives) one obtains that \citep{DKSZ1996}\ %
\begin{equation} \label{KineticEnergy2}
K=-\frac{1}{2}\int\limits^\infty_{-\infty}\psi\hat {\mathcal
H}\psi_u\D u.
\end{equation}

\section{Hamiltonian formalism}
\label{sec:Hamiltonian formalizm}

Conformal mapping makes possible an extension of the Hamiltonian formalism of Eqs. \e{Hamiltpsieta} for single-valued function $\eta$ of $x$ into a general  multi-valued case, i.e. to the parametrization   \e{xyu}. For that we notice that the Hamiltonian Eqs. \e{Hamiltpsieta} can be obtained from the minimization of the action functional%
\begin{equation} \label{ActionS}
S=\int L \D t
\end{equation}
with the Lagrangian%
\begin{equation} \label{Lagrangiandef}
L=\int\limits^\infty_{-\infty}\psi\eta_t\D x-H.
\end{equation}
We now generalize the Lagrangian \e{Lagrangiandef} into multi-valued $\eta$ through the parametrization   \e{xyu} as
\begin{equation} \label{Lagrangiandef2}
L=\int\limits^\infty_{-\infty}\psi(y_t x_u-x_t y_u)\D u-H
\end{equation}
with the Hamiltonian %
\begin{equation} \label{Hwithx}
H=-\frac{1}{2}\int\limits^\infty_{-\infty}\psi\hat {\mathcal
H}\psi_u\D u+\frac{g}{2}\int \limits_{-\infty}^{\infty} y^2\,x_u\D  u+\alpha\int\limits^{\infty}_{-\infty}\left
(\sqrt{x_u^2+y_u^2}-x_u\right )\D u.
\end{equation}
as follows from  Eqs. \e{Hamilttotal}, \e{PotentialEnergy2}
and \e{KineticEnergy2}. Here we used the change of variables in  $\eta_t\D x$  of Eq.  \e{Lagrangiandef} from $(x,t)$ into $(u,t)$ which results in   $\eta_tdx=(y_t x_u-x_t y_u)\D u$
  (see also Appendix A.2 of Ref. \cite{DyachenkoLushnikovKorotkevichPartIStudApplMath2016} for more details).

Using  Eq. \e{xHy} to explicitly express $x(u,t)$ as the functional of $y(u,t),  $ one can rewrite the Hamiltonian $H$   \e{Hwithx} as follows
\begin{equation} \label{Hpsiyonly}
H=-\frac{1}{2}\int\limits^\infty_{-\infty}\psi\hat {\mathcal
H}\psi_u\D u+\frac{g}{2}\int \limits_{-\infty}^{\infty} y^2\,(1-\hat
{\mathcal H} y_u)\D  u+\alpha\int\limits^{\infty}_{-\infty}\left
(\sqrt{(1-\hat {\mathcal H} y_u)^2+y_u^2}-1+\hat {\mathcal H}
y_u\right )\D u.
\end{equation}
 We can use either Eq.  \e{Hwithx} or \e{Hpsiyonly} at our convenience for finding the dynamic equations.

Vanishing of a variation $\delta S=0$ of Eq. \e{ActionS} over $\psi$ together with Eq. \e{Lagrangiandef2}
results in
\begin{equation} \label{psivariation}
y_t x_u-x_t y_u=-\hat {\mathcal H}\psi_u=\frac{\delta H}{\delta
\psi}
\end{equation}
which gives kinematic BC \e{fullconformal1}.

Variations over $x$ and $y$  must satisfy the condition \e{xHy}. To ensure that condition we introduce the modification $\tilde L$ of the Lagrangian \e{Lagrangiandef2} and the modified action $\tilde S$ by adding the term with the Lagrange multiplier
 $f(u,t)$ as %
\begin{equation} \label{Ltilde}
\tilde L=L+\int\limits^\infty_{-\infty}f[y-\hat {\mathcal H}
(x-u)]\D u, \quad \tilde S=\int \tilde L\D t,
\end{equation}
which does not change Eq.
\e{psivariation}.

To ensure the most compact derivation of the dynamical equations from the variation of $\tilde S$, we use the Hamiltonian \e{Hpsiyonly}  (which does not contain $x$) while we keep $x$ (not expressing it as a functional of $y$) in the remaining terms of the modified action $\tilde S$ beyond $H$. Then a vanishing of a variation  $\delta \tilde S=0$ over $x$ and $y,$ together with \e{Lagrangiandef2},  \e{Hwithx}  and \e{Ltilde},
result in  Eqs.
\begin{align}
&y_u\psi_t - y_t\psi_u + \hat {\mathcal H} f =0 \label{Var_by_x}
\end{align}
and
\begin{align}
-&x_u\psi_t + x_t\psi_u+f= \frac{\delta H}{\delta y}=g yx_u-g\hat
{\mathcal H}(yy_u)-\alpha\hat {\mathcal H} \frac{\p  }{\p
u}\frac{x_u}{|z_u|}-\alpha\frac{\p  }{\p u}\frac{y_u}{|z_u|}
 ,\label{Var_by_y}
\end{align}
respectively.   Here we used that
\begin{align}\label{Fxvar}
\frac{\delta  F(x-u)}{\delta y}=\hat {\mathcal H}\frac{\delta
F(x-u)}{\delta x}
\end{align} for any functional $F$ of $x(u)-u=-\hat {\mathcal H}y$.%,  see Appendix %???
%for more details.

Excluding the Lagrange multiplier $f$ from Eqs. \e{Var_by_x} and
\e{Var_by_y} by applying $\hat {\mathcal H}$ to Eq.  \e{Var_by_y}
and subtracting the result from Eq.  \e{Var_by_x} we recover Eq.
\e{fullconformal2}.

We note that there are two alternatives to using Eqs.
\e{Var_by_x} and \e{Var_by_y}. First one is to  keep $x$ in the
Hamiltonian $H$   \e{Hamilttotal}, \e{PotentialEnergy2},
\e{KineticEnergy2}  (instead of replacing it by $u-\hat {\mathcal
H}(x-u)  $ as was done in Eq. \e{Hpsiyonly}). Then vanishing
variations of $\tilde S$ \e{Ltilde} over $x$ or $y$ results in
modification of  Eqs. \e{Var_by_x} and \e{Var_by_y}.   Excluding $f$
from these modified Eqs. still results in Eq. \e{fullconformal2} as
was obtained in Ref. \cite{DKSZ1996}.
 Second alternative is to  replace $x$ by $u-\hat {\mathcal H}(x-u)  $ in Eqs.  \e{ActionS}, \e{Lagrangiandef2}
 and use the Hamiltonian \e{Hpsiyonly}.     Then a vanishing variation of $ S$ \e{ActionS} over $y$ results in   \begin{equation}\label{fullconformal3}
\hat {\mathcal H}(\psi_t y_u - \psi_u y_t)-\psi_t x_u + \psi_ux_t
=\frac{\delta H}{\delta y}=g yx_u-g\hat {\mathcal
H}(yy_u)-\alpha\hat {\mathcal H} \frac{\p  }{\p
u}\frac{x_u}{|z_u|}-\alpha\frac{\p  }{\p u}\frac{y_u}{|z_u|} .
\end{equation}
Applying $-\hat {\mathcal H}$ to Eq. \e{fullconformal3} we again
recover Eq. \e{fullconformal2}. A variational derivative of the
Hamiltonian over $\psi$ in all  cases is  given by Eq.
\e{psivariation}.

 The second alternative allows to  obtain Eq.
\e{fullconformal2} without the use of the Lagrange multiplier $f$.
Below we  use Eqs. \e{Var_by_x} and \e{Var_by_y} because they allow
to significantly simplify subsequent transformations.

Applying $-\hat {\mathcal H}$ to Eq. \e{Var_by_x} and adding it to
Eq. \e{Var_by_y} recovers Eq. \e{fullconformal3}. We  use Eqs.
\e{psivariation}
 and \e{fullconformal3} to rewrite   Eqs. \e{fullconformal1} and \e{fullconformal2}  in the ``symplectic"  Hamiltonian form \cite{ZD2012}

\begin{equation} \label{symplecticlike1}
\hat \Omega{\bf Q}_t=\frac{\delta H}{\delta {\bf Q}}, \quad \quad {\bf Q}\equiv\begin{pmatrix}y \\
\psi
\end{pmatrix},
\end{equation}
where the symplectic operator $\hat \Omega$ is given by%
\begin{equation} \label{Omegadef}
\hat \Omega=\begin{pmatrix}\hat \Omega_{11} & \hat \Omega_{12} \\
\hat \Omega_{21} & 0 \\
\end{pmatrix},
\end{equation}
which is $2\times 2$ skew-symmetric matrix operator with %
\begin{equation} \label{Omega12skew}
\hat \Omega_{21}^\dagger=-\hat \Omega_{12},
\end{equation}
 Here $\hat \Omega_{21}^\dagger $is the adjoint operator, $\langle f,\hat \Omega_{ij} g\rangle\equiv\langle \hat \Omega_{ij}^\dagger f,g\rangle$, $\ i,j=1,2$,  with respect to the  scalar product $\langle f,g\rangle=\int^{\infty}_{-\infty}f(u)g(u)du$. Also $\hat \Omega_{11}$ is the skew-symmetric operator%
\begin{equation} \label{Omega11skew}
\hat \Omega_{11}^\dagger= -\hat \Omega_{11}.
\end{equation}
Eqs.  \e{symplecticlike1} and \e{Omegadef} expressed in components are given by
\begin{equation}\label{psiysymplectic}
\begin{split}
& \hat \Omega_{11}y_t+\hat \Omega_{12}\psi_t=\frac{\delta H}{\delta y}, \\
& -\hat \Omega_{12}^\dagger y_t=\frac{\delta H}{\delta \psi}.
\end{split}
\end{equation}
Using Eqs.   \e{psivariation} and \e{fullconformal3} we obtain that  %
\begin{equation} \label{Omega12}
\hat \Omega_{21}q=x_uq+y_u\hat {\mathcal H}q=(1-\hat {\mathcal
H}y_u)q+y_u\hat {\mathcal H}q
\end{equation}
for any function $q=q(u).$ Using Eqs. \e{xHy} and \e{fullconformal3}   we obtain that%
\begin{equation} \label{Omega11}
\hat \Omega_{11}q=-\hat {\mathcal H}(\psi_uq)-\psi_u\hat {\mathcal
H}q, \quad \Omega_{12}q=-x_uq+\hat {\mathcal H}(y_uq)=-(1-\hat
{\mathcal H}y_u)q+\hat {\mathcal H}(y_uq).
\end{equation}
Using integration by parts and definition \e{Hilbertdef} in Eqs. \e{Omega12}, \e{Omega11} ensures a validity of Eqs. \e{Omega12skew} and  \e{Omega11skew}.
We note
that Eqs. \e{symplecticlike1}-\e{Omega11} are valid for any Hamiltonian, not only for the Hamiltonian  \e{Hpsiyonly} provided we  derive them from the variation of action \e{Ltilde}.
Because Eqs.
 \e{psivariation} and \e{fullconformal3}
are obtained directly from the variation principle, the symplectic form, corresponding to the symplectic operator $\hat \Omega$ \e{Omegadef}, is closed and nondegenerate (see Ref. \cite{ArnoldClassicalMechanics1989}).

Eqs. \e{psivariation} and \e{fullconformal3} are not resolved with respect to the time derivatives $y_t$ and $\psi_t$.
It is remarkable that the symplectic operator $\hat \Omega$ \e{Omegadef} can be explicitly inverted. We first find the explicit expression for $y_t$ using Eq. \e{psivariation}
rewritten in the complex form %
\begin{equation} \label{zKinematic}
z_t\bar z_u-\bar z_tz_u=-2\I\hat {\mathcal H}\psi_u=2\I\frac{\delta
H}{\delta \psi},
\end{equation}
where $\bar f(w)$ means a complex conjugate of a function $f(w)$.
Note that the complex conjugation $\bar f(w)$ of $f(w)$  in this paper is understood as applied with the assumption that $f(w)$ is the complex-valued function of the real argument $w$ even if $w$ takes the complex values so that %
\begin{equation} \label{bardef}
 \bar f(w)\equiv \overline {f(\bar{w})}.
\end{equation}
That definition ensures
  the analytical continuation of $f(w)$ from
the real axis  $w=u$ into the complex plane of $w\in\mathbb{C.}$

We use the Jacobian %
\begin{equation} \label{Jdef}
J=x_u^2+y_u^2=z_u\bar z_u=|z_u|^2
\end{equation}
which is nonzero for $w\in\C^{-}$
because $z=z(w,t)$ is the conformal mapping there. Dividing Eq. \e{zKinematic} by $J$ we obtain that%
\begin{equation} \label{ztzbart}
\frac{z_t}{z_u}-\frac{\bar z_t}{\bar z_u}=-\frac{2\I}{J}\hat
{\mathcal H}\psi_u=\frac{2\I}{J}\frac{\delta H}{\delta \psi}.
\end{equation}
Here $\frac{z_t}{z_u}$ is analytic in $\C^-$ and $\frac{\bar z_t}{\bar z_u}$ is analytic in $\C^+$.

It is convenient to introduce the  operators
\begin{equation} \label{Projectordef}
\hat P^-=\frac{1}{2}(1+\I \hat {\mathcal H})  \quad\text{and}\quad
\hat P^+=\frac{1}{2}(1-\I \hat {\mathcal H})
\end{equation}
which are the projector operators of a  function $q(u)$ defined at the real
line $w=u$ into  functions $q^+(u)$ and $q^-(u)$ analytic in $w\in\mathbb{C}^-$ and
$w\in\mathbb{C}^+$, respectively, such that %
\begin{equation} \label{qprojectiondef}
q=q^++q^-.
\end{equation}
Here we assume that $q(u)\to 0$ for $u\to \pm\infty$.
Eqs.    \e{Projectordef} imply that
\begin{equation} \label{Pfm}
\hat P^+(q^++q^-)=q^+ \quad\text{and}\quad  \hat P^-(q^++q^-)=q^-,
\end{equation}
see more discussion of the operators    \e{Projectordef}  in
Appendix \ref{sec:AppA}. %  \ref{sec:Projectors}.
 Also notice that Eqs.     \e{Projectordef} result in
the identities%
\begin{equation} \label{Hfpm}
\hat {\mathcal H} q=\I[q^+-q^-]
\end{equation}
and
\begin{equation} \label{Poperidentitties}
\hat P^++\hat P^-=1, \ (\hat P^+)^2= \hat P^+, \ (\hat P^-)^2=\hat P^-, \\ \  \hat P^+\hat P^-=\hat P^-\hat P^+=0.
\end{equation}

Applying $\hat P^-$ to Eq. \e{ztzbart} and multiplying by $z_u$ after that we find that
\begin{equation} \label{ztexplicit}
z_t=-z_u\hat P^-\left [\frac{2\I}{J}\hat {\mathcal H}\psi_u\right
]=z_u\hat P^-\left [\frac{2\I}{J}\frac{\delta H}{\delta \psi}\right
]
\end{equation}
which is explicit solution for time derivative in complex form. Taking the real and   imaginary parts we obtain that
\begin{equation} \label{ytexplicit}
y_t=(y_u\hat {\mathcal H}-x_u)\left [\frac{1}{J}\hat {\mathcal
H}\psi_u\right ]=-(y_u\hat {\mathcal H}-x_u)\left
[\frac{2}{J}\frac{\delta H}{\delta \psi}\right ]
\end{equation}
and
\begin{equation} \label{xtexplicit}
x_t=(x_u\hat {\mathcal H}+y_u)\left [\frac{1}{J}\hat {\mathcal
H}\psi_u\right ]=-(x_u\hat {\mathcal H}+y_u)\left
[\frac{2}{J}\frac{\delta H}{\delta \psi}\right ].
\end{equation}

We now multiply Eq. \e{Var_by_x} by $x_u$ and add to Eq. \e{Var_by_y} multiplied by $y_u$ to exclude $\psi_t$ which results in %
\begin{equation} \label{fHf}
\psi_u(y_tx_u-y_ux_t)+y_u\frac{\delta H}{\delta y}=x_u\hat {\mathcal
H}f+y_uf=-\I z_u\hat P^-f+\I \bar z_u\hat P^+f.
\end{equation}
We use Eq. \e{psivariation} in l.h.s. of Eq. \e{fHf} to exclude time derivative and apply $P^-$ to it  to obtain Eq.%
\begin{equation} \label{fHf2}
\hat P^-f=\frac{\I}{z_u}\hat P^-\left [y_u\frac{\delta H}{\delta
y}-\psi_u\hat {\mathcal H} \psi_u  \right ]= \frac{\I}{z_u}\hat
P^-\left [y_u\frac{\delta H}{\delta y}+\psi_u\frac{\delta H}{\delta
\psi}  \right ],
\end{equation}
which does not contain any time derivative.
Taking a sum of Eq.  \e{Var_by_x} multiplied by $\I$ and Eq. \e{Var_by_y} result in%
\begin{equation} \label{varsum}
\psi_t\bar z_u-\bar z_t\psi_u-2\hat P^-f=-\frac{\delta H}{\delta y}.
\end{equation}
Excluding $\hat P^-f$ and $\bar z_t$ in Eq. \e{varsum} through  Eqs. \e{fHf2} and \e{ztexplicit} we obtain%
\begin{equation} \label{psiexpl}
\psi_t=-\psi_u\hat P^+\left [\frac{2\I}{J}\frac{\delta H}{\delta \psi}\right ]+\frac{2\I}{J}\hat P^-\left [y_u\frac{\delta H}{\delta y}+\psi_u\frac{\delta H}{\delta \psi}  \right ]-\frac{1}{\bar z_u}\frac{\delta H}{\delta y}.
\end{equation}
Using Eq. \e{Pfm} we transform Eq. \e{psiexpl} into
\begin{equation} \label{psiexpl2}
\psi_t=-\psi_u\hat {\mathcal H}\left [\frac{1}{J}\frac{\delta
H}{\delta \psi}\right ]-\frac{1}{J}\hat {\mathcal H}\left
[\psi_u\frac{\delta H}{\delta \psi}  \right
]-\frac{x_u}{J}\frac{\delta H}{\delta y}-\frac{1}{J}\hat {\mathcal
H}\left [y_u\frac{\delta H}{\delta y}  \right ].
\end{equation}

Eqs. \e{ytexplicit} and \e{psiexpl2} can be written in the general Hamiltonian form

\begin{equation} \label{implectic2}
 {\bf Q}_t=\hat {\mathcal R}\frac{\delta H}{\delta {\bf Q}}, \quad {\bf Q}\equiv\begin{pmatrix}y \\
\psi
\end{pmatrix},
\end{equation}
where%
\begin{equation} \label{Rdef}
\hat {\mathcal R}=\hat \Omega^{-1}=\begin{pmatrix}0 & \hat {\mathcal R}_{12} \\
\hat {\mathcal R}_{21} &  \hat {\mathcal R}_{22} \\
\end{pmatrix}
\end{equation}
is $2\times 2$ skew-symmetric matrix operator with the components
\begin{equation}\label{Rmatrdef}
\begin{split}
& \hat {\mathcal R}_{11}q=0,\\
&\hat {\mathcal R}_{12}q=\frac{x_u}{J}q-y_u\hat {\mathcal H}\left ( \frac{q}{J}\right ), \\
& \hat {\mathcal R}_{21}q=-\frac{x_u}{J}q-\frac{1}{J}\hat {\mathcal H}\left (y_u q\right ), \quad  \hat {\mathcal R}_{21}^\dagger=-\hat {\mathcal R}_{12},\\
 & \hat {\mathcal R}_{22}q=-\psi_u\hat {\mathcal H}\left ( \frac{q}{J}\right ) -\frac{1}{J}\hat {\mathcal H}\left (\psi_u q\right ),\quad  \hat {\mathcal R}_{11}^\dagger=-\hat {\mathcal R}_{11}.
\end{split}
\end{equation}
We call $\hat {\mathcal R}=\hat \Omega^{-1}$ by the ``implectic"
operator (sometimes such type of inverse of the symplectic operator
is also called by the co-symplectic operator, see e.g. Ref.
\cite{WeinsteinJDiffGeom1983,MorrisonRevModPhys1998}).

Writing Eq. \e{implectic2} in components we also obtain that
\begin{equation}\label{Rsimplectic}
\begin{split}
& y_t=\hat {\mathcal R}_{12}\frac{\delta H}{\delta \psi}, \\
& \psi_t=\hat {\mathcal R}_{21}\frac{\delta H}{\delta y} +\hat
{\mathcal R}_{22}\frac{\delta H}{\delta\psi}.
\end{split}
\end{equation}

Comparing Eqs. \e{symplecticlike1} and \e{implectic2} we conclude
that $\hat {\mathcal R}=\hat \Omega^{-1}$ which can be  confirmed
by the direct calculation that
\begin{equation} \label{ROmegaI}
\hat {\mathcal R}\hat \Omega= \hat \Omega\hat {\mathcal R}=I,
\end{equation}
 where $I$ is the identity operator.% See Appendix \ref{sec:AppA} for example of such calculation ???.

We use Eqs. \e{implectic2}
and \e{Rdef} to define the Poisson  bracket
\begin{equation}\label{PoissonBracketsRDef}
\begin{split}
& \{F,G\}=\sum\limits^{2}_{i,j=1}\int \limits^{\infty}_{-\infty}\D
u\left ( \frac{\delta F}{\delta Q_i }\hat {\mathcal R}_{ij}
\frac{\delta G}{\delta Q_j } \right )=\int
\limits^{\infty}_{-\infty}\D u\left ( \frac{\delta F}{\delta y }\hat
{\mathcal R}_{12}  \frac{\delta G}{\delta \psi } +\frac{\delta
F}{\delta\psi }\hat {\mathcal R}_{21}  \frac{\delta G}{\delta y
}+\frac{\delta F}{\delta \psi }\hat {\mathcal R}_{22}  \frac{\delta
G}{\delta \psi }\right )
\end{split}
\end{equation}
between arbitrary functionals $F$ and $G$ of $\bf Q$.    It is clear
from that definition that  any functionals $\xi$ and $\eta$ of $y$
only commute to each other, i.e.  $\{\xi,\eta\}=0.$

Eq. \e{PoissonBracketsRDef} allows to  rewrite Eqs.  \e{implectic2}
and \e{Rdef} in the non-canonical Hamiltonian form corresponding to Poisson mechanics as follows%
\begin{equation}\label{hamiltoniancanonicalpoisson}
\begin{split}
 {\bf Q}_t=\{{\bf  Q},H\}.
\end{split}
\end{equation}
The Poisson bracket requires to satisfy a Jacobi identity
\begin{align} \label{JacobiIdentity}
\{F,\{G,L\}\}+\{G,\{L,F\}\}+\{L,\{F,G\}\}=0
\end{align}for arbitrary functionals $F$, $G$ and $L$ of $\bf Q$. The Jacobi identity is ensured by our use of the variational principle for the action \e{Ltilde}.
 % The Jacobi identity can be also verified  directly from Eqs. \e{Rmatrdef} and \e{PoissonBracketsRDef}
%(see Appendix ???).

A functional $F$ is the constant of motion of Eq. \e{hamiltoniancanonicalpoisson}
provided $\{{ F},H\}=0.$
It  follows from Eq. \e{PoissonBracketsRDef} that any functionals $F$ and $G,$ which depend only on $y$,  commute with each other, i.e.  $\{{ F},G\}=0.$ We note that the derivation of  Eqs. \e{implectic2}-\e{hamiltoniancanonicalpoisson}
%{Rmatrdef}
is valid for any Hamiltonian, not only for the Hamiltonian
\e{Hpsiyonly}, because we  derive these equations starting from the
variation of action \e{Ltilde}. It implies that Eq.
\e{PoissonBracketsRDef} has no Casimir invariant (the constant of
motion which does not depend on the particular choice of the
Hamiltonian $H$, see e.g. Refs.
\cite{WeinsteinJDiffGeom1983,ZakharovKuznetsovUspekhiFizNauk1997}).
Beyond our standard Hamiltonian \e{Hpsiyonly}, one can also apply
Eqs. \e{PoissonBracketsRDef}, \e{hamiltoniancanonicalpoisson} to
more general cases as discussed in Section
\ref{ref:Generalizedhydrodynamicsintegrability}.

\section{Dynamic equations for the Hamiltonian
\e{Hpsiyonly}}
\label{Dynamicalequations}

Eq. \e{ytexplicit}  provides the kinematic BC solved to $y_t.$ Eq.  \e{psiexpl2} with the Hamiltonian \e{Hpsiyonly} can be simplified as follows.  We first notice that using Eq. \e{Omega11},  the gravity part of the variational derivative \e{fullconformal3} can be represented as follows
  \begin{equation}\label{fullconformal3g}
\left .\frac{\delta H}{\delta y}\right |_{\alpha=0}=g yx_u-g\hat
{\mathcal H}(yy_u)=-g\hat \Omega_{12}y.
\end{equation}
Then the contribution of that gravity part into r.h.s. of Eq. \e{Omega11} is given by
  \begin{equation}\label{fullconformal3g}
\hat {\mathcal R}_{21}\left .\frac{\delta H}{\delta y}\right
|_{\alpha=0}=-g\hat {\mathcal R}_{21}\hat \Omega_{12}y=-gy,
\end{equation}
where we use the definition \e{Rmatrdef} and Eq. \e{ROmegaI}.

Second step is to simplify the surface tension part
\begin{equation}\label{fullconformal3alpha} \left .\frac{\delta
H}{\delta y}\right |_{g=0}=-\alpha\hat {\mathcal H} \frac{\p  }{\p
u}\frac{x_u}{|z_u|}-\alpha\frac{\p  }{\p u}\frac{y_u}{|z_u|}
\end{equation}
 of the variational derivative \e{fullconformal3}. We also notice the identity
\begin{equation} \label{xuyuidentit}
x_u\frac{\p  }{\p u}\frac{x_u}{|z_u|}+y_u\frac{\p  }{\p u}\frac{y_u}{|z_u|}=\frac{1}{2|z_u|}\frac{\p  }{\p u}(x^2_u+y_u^2)+(x^2_u+y_u^2)\frac{\p  }{\p u}\frac{1}{|z_u|}\\ =0
\end{equation}
which is the particular case  of the identity $$
 \frac{\delta F}{\delta x} x_u+\frac{\delta F}{\delta y} y_u  \equiv0$$for  general parametrization invariant
functionals $F((x(u), y(u))$, see e.g. Refs. \cite{MorrisonPhysPlasm2005,FlierlMorrisonSwaminathanArxiv2018}.
 Eq.  \e{xuyuidentit} corresponds to $F= \int ^\infty_{-\infty}(|z_u|-x_u)\D u$ which is the parametrization invariant functional because it represents the arclength of the surface (minus the arclength of unperturbed surface) and thus is independent on the particular surface parametrization $(x(u), y(u))$, see also Eq. \e{PotentialEnergy} and discussion after it.  %and the similar identity
%
%\begin{equation} \label{zuzbaruidentit}
%z_u\frac{\p  }{\p u}\frac{\bar z_u}{|z_u|}+\bar z_u\frac{\p  }{\p u}\frac{z_u}{|z_u|}=0.
%\end{equation}
%

The contribution of the surface tension part into r.h.s. of Eq. \e{Omega11} is given by
  \begin{equation}\label{fullconformal3alpha}
\hat {\mathcal R}_{21}\left .\frac{\delta H}{\delta y}\right
|_{g=0}=\hat {\mathcal R}_{21} \left [-\alpha\hat {\mathcal H}
\frac{\p  }{\p u}\frac{x_u}{|z_u|}-\alpha\frac{\p  }{\p
u}\frac{y_u}{|z_u|} \right ]=\frac{\alpha }{x_u}\frac{\p  }{\p
u}\frac{y_u}{|z_u|},
\end{equation}
where we used Eqs. \e{Rmatrdef}, \e{ROmegaI} and expressed $\frac{\p  }{\p u}\frac{x_u}{|z_u|}$ through the identity \e{xuyuidentit}.  Eq. \e{fullconformal3alpha} has a removable singularity  at $x_u=0$. To explicitly remove that singularity we perform the explicit differentiation in r.h.s of this Eq. to obtain that
  \begin{equation}\label{fullconformal3alphanoxu}
\hat {\mathcal R}_{21}\left .\frac{\delta H}{\delta y}\right
|_{g=0}=\frac{\alpha(x_uy_{uu}-x_{uu}y_u) }{|z_u|^3},
\end{equation}
which provides the expression for the pressure jump  \e{Palpha}.
Using Eqs. \e{psivariation},\e{fullconformal3g} and \e{fullconformal3alphanoxu} we obtain a particular form of Eq. \e{psiexpl2} for the Hamiltonian \e{Hpsiyonly} as follows
\begin{equation} \label{psiexpl2alpha}
\psi_t=\psi_u\hat {\mathcal H}\left [\frac{1}{|z_u|^2}\hat {\mathcal
H}\psi_u\right ]+\frac{1}{|z_u|^2}\hat {\mathcal H}\left [\psi_u\hat
{\mathcal H} \psi_u \right ]-gy+\frac{\alpha(x_uy_{uu}-x_{uu}y_u)
}{|z_u|^3}.
\end{equation}
Eqs.  \e{xHy}, \e{ytexplicit} and \e{psiexpl2alpha} form a closed set of equations defined on the real line $w=u.$ That system was first obtained in Ref. \cite{DKSZ1996}
with the surface tension term in the form \e{fullconformal3alpha}.
We notice that the same system can be obtained directly from Eqs. \e{xyu},\e{xyinfinity},\e{laplace}-\e{ndef},\e{dynamic1},\e{Palpha} and the definition of the conformal mapping \e{zwdef} without any use of the variational principle of Section \e{sec:Hamiltonian formalizm}. However, such alternative derivation is significantly more cumbersome.

\section{Dynamic equations in the complex form} \label{DynamicalequationsComplex}

Dynamical Eqs.  \e{psiexpl2} are defined on the real line $w=u$ with
the analyticity of $z(w,t)$ and $\Pi(w,t)$ in $w\in \C^-$  taken
into account through the Hilbert operator $\hat {\mathcal H}.$ For
the analysis of surface hydrodynamics, it is efficient to consider
the analytical continuation of  $z(w,t)$ and $\Pi(w,t)$ into $w\in
\C^+$ with the time-dependent complex singularities of these
functions fully determine their properties. The projector operators
\e{Projectordef}
are convenient tools for such analytical continuation with%
\begin{equation} \label{PiP}
\Pi=\psi+\I \hat {\mathcal H}\psi=2\hat P^-\psi
\end{equation}
and
\begin{equation} \label{zP}
z-u=-\hat {\mathcal H}y+\I y=2\I\hat P^-y,
\end{equation}
see Appendix \ref{sec:AppA} for more details. Analytical
continuation of Eqs. \e{PiP} and \e{zP} into complex plane $w\in\C$
amounts to a straightforward replacing $u$ by $w$ in Eq.  \e{Puminus}
(as well as in Eqs. \e{fplus} and \e{fminus}, see also Appendix
\ref{sec:AppA}) which is always allowed provided $w\in\C^+$ and
$w\in\C^-$ for $\hat P^+q(w)$ and $\hat P^-q(w)$, respectively. This
is possible because  the pole singularity  at $u'=u\pm\I 0$ in the
integrand of Eq. \e{Puminus} does not cross the integration contour
$-\infty<u'<\infty$ as $w$ continuously  changes from $w=u$ into the
complex values. Analytical continuation in the opposite direction
(i.e. into   $w\in\C^+$ for $\hat P^-q(w)$ and  $w\in\C^-$ for $\hat
P^+q(w)$)  however requires to move/deform the  integration contour
$-\infty<u'<\infty$  which is possible only so long as complex
singularities are not reached. We also remind our definition
\e{bardef} of complex conjugation which ensures how to define $\bar
f(w)$ for $w\in\C$. Another convenient way of analytical
continuation from the real line $w=u$ into $\C$ is to use Eqs.
\e{fpm0}-\e{fminus}.
 However, such continuation into   $w\in\C^+$ for $\hat P^-q(w)$ and  $w\in\C^-$ for $\hat P^+q(w)$)  is limited by the convergence of integrals in Eqs. \e{fplus} and \e{fminus} which implies that $|Im(w)|$ cannot exceed   the distance of  a singularity closest to the real axis. We also note that if the function $q(w)$ is analytic in $\C^-$ then    $\bar{q}({w})$
 is analytic in $\C^+$ and vise versa.

We replace variations over  $y$ and $\psi$ of Section \ref{sec:Hamiltonian formalizm} by variation over $z$, $\bar z$, $\Pi$ and $\bar \Pi$ according to %
\begin{equation} \label{varcomplex}
\frac{\delta}{\delta y}=2\I\hat P^+\frac{\delta}{\delta z}-2\I\hat P^-\frac{\delta}{\delta \bar z} \quad \text{and} \quad \frac{\delta}{\delta \psi}=2\hat P^+\frac{\delta}{\delta \Pi}+2\hat P^-\frac{\delta}{\delta \bar \Pi}
\end{equation}
as follows from Eqs. \e{PiP} and \e{zP}, see also Eq. \e{Fxvar}. % and Appendix ???.
Here we used that %
\begin{equation} \label{zPi}
x=\frac{z+\bar z}{2}, \ y=\frac{z-\bar z}{2\I} \ \text{and} \ \psi=\frac{\Pi+\bar \Pi}{2},
\end{equation}
as follows from Eqs. \e{zwdef} and \e{ComplexPotentialdef}.
In variational derivatives \e{varcomplex} we assume that $z$, $\bar z$, $\Pi$ and $\bar \Pi$ are independent variables. %The Poisson bracket in these variables is provided in Appendix ???

Applying $\hat P^-$ to Eqs. \e{ztexplicit} and \e{psiexpl} together with Eqs. \e{varcomplex} and \e{zPi}, we obtain the following dynamic equations
\begin{align}\label{Zt}
&z_t=\I Uz_u, \\
\label{Pit}
&\Pi_t=\I U\Pi_u-B-\mathcal{P,}
\end{align}
where %
\begin{align} \label{Udef1}
U=4 \hat P^-\left  \{\frac{1}{J}\left [\hat P^-\frac{\delta H}{\delta \bar \Pi}+\hat P^+\frac{\delta H}{\delta  \Pi} \right  ] \right \},
\end{align}
is the complex transport velocity,
\begin{equation} \label{Pintdef}
\mathcal{P}=-4\I \hat P^-\left  \{\frac{1}{J}\left [\hat P^-\left (\bar z_u\frac{\delta H}{\delta \bar z}\right )-\hat P^+\left (z_u\frac{\delta H}{\delta z}\right ) \right  ] \right \},
\end{equation}
and
\begin{equation} \label{Bintdef}
B=-4\I \hat P^-\left  \{\frac{1}{J}\left [\hat P^-\left (\bar \Pi_u\frac{\delta H}{\delta \bar \Pi}\right )-\hat P^+\left (\Pi_u\frac{\delta H}{\delta \Pi}\right ) \right  ] \right \}.
\end{equation}
Here we used that $z$ and $\Pi$ and analytic in $\C^-$ while $\bar z$ and $\bar \Pi$ are analytic in $\C^+.$ Taking an imaginary part of Eq. \e{Zt} and a real  part of Eq. \e{Pit} one can recover Eqs. \e{ytexplicit} and \e{psiexpl2}.

Eqs. \e{Zt}-\e{Bintdef}  are convenient for analytical study.
A version of dynamic equations is obtained by the  change of variables (suggested in Ref.  \citet{Dyachenko2001})%
\begin{align} \label{RVvar1}
&R=\frac{1}{z_u}, \\ \label{RVvar2}
&V=\I\frac{\p \Pi}{\p z}=\I R \Pi_u.
\end{align}
Eqs. \e{Zt} and \e{Pit} in terms of variables \e{RVvar1} and \e{RVvar2} take the following form
 \begin{align}
\frac{\partial R}{\partial t} &= \I \left(U R_u - R U_u \right), \label{Reqn0}\\
\frac{\partial V}{\partial t} &= \I \left[U V_u - R (B_u+\mathcal{P}_u \right) ]. \label{Veqn0}
\end{align}
These dynamic equations are valid for any Hamiltonian.
They are also convenient for numerical simulations to avoid a numerical instability at small spatial scales, see e.g. Ref. \cite{ZakharovDyachenkoProkofievEuropJMechB2006} and related analysis of weakly nonlinear case in Ref. \cite{LushnikovZakharov2005}. Note that $R$ and $V$ include only a derivative of the conformal mapping \e{zwdef} and the complex potential $\Pi$ over $w$ while $z(w,t)$ and $\Pi(w,t)$ are recovered from solution of these Eqs. as $z=\int \frac{1}{R} dw$ and $\Pi=-\I\int \frac{V}{R}dw$. Respectively, these relation can be used to recover the integrals of motion \e{Pxdef2}, \e{Pydef2} and \e{Mass2}
from $R$ and $V.$

We now rewrite  our standard Hamiltonian \e{Hpsiyonly} in terms of variables  $z$, $\bar z$, $\Pi$ and $\bar \Pi$  which gives that%
\begin{equation} \label{HamiltzPi}
H=\int\limits^\infty_{-\infty}du\left [ \frac{\I}{8}(\Pi_u\bar\Pi-\Pi\bar \Pi_u)-\frac{g}{16}(z-\bar z)^2(z_u+\bar z_u)+\alpha\left (\sqrt{z_u\bar z_u}-\frac{z_u+\bar z_u}{2}\right )\right ].
\end{equation}
Eqs. \e{Udef1}-\e{Bintdef} and \e{HamiltzPi}
results in
\begin{align} \label{Udef2}
U=\I \hat P^-\left  \{\frac{1}{J}\left[\Pi_u-\bar \Pi_u\right  ] \right \}=\hat P^-(R\bar V+\bar R V),
\end{align}
\begin{equation} \label{Pintdef2}
\mathcal{P}=-\I g(z-w)-2\I\alpha  \hat P^-(Q_u\bar Q-Q\bar Q_u ),
\end{equation}
and
\begin{equation} \label{Bintdef2}
B= \hat P^-\left  \{\frac{|\Pi_u|^2}{|z_u|^2} \right \}=\hat P^-(|V|^2),
\end{equation}
where %
\begin{equation} \label{Qdef}
Q\equiv\frac{1}{\sqrt{z_u}}=\sqrt{R}.
\end{equation}

Plugging in Eqs. \e{Udef2}-\e{Bintdef2} into Eqs. \e{Reqn0} and \e{Veqn0} we obtain

 \begin{align}
\frac{\partial R}{\partial t} &= \I \left(U R_u - R U_u \right), \label{Reqn}\\
\frac{\partial V}{\partial t} &= \I \left[ U V_u - R B_u \right ]+ g(R-1)-2\alpha R  \hat P^-\frac{\p }{\p u}(Q_u\bar Q-Q\bar Q_u ). \label{Veqn}
\end{align}
Other authors have referred to these equations as the ``Dyachenko''
equations (\citet{Dyachenko2001}) which serve as a basis for
numerical study of free surface hydrodynamics. They can be also
immediately rewritten fully in terms of $Q$ and $V$ as follows
 \begin{align}
&\frac{\partial Q}{\partial t} = \I \left(U Q_u -\frac{1}{2} Q U_u \right), \label{ReqnQ}\\
&\qquad U=\hat P(Q^2\bar V+\bar Q^2V), \quad  \label{qdef3}\\
&\frac{\partial V}{\partial t} = \I \left[ U V_u - Q^2 \hat P^- \frac{\p}{\p u}(|V|^2)\right ]+ g(Q^2-1)-2\alpha Q^2  \hat P^-\frac{\p }{\p u}(Q_u\bar Q-Q\bar Q_u ). \label{VeqnQ}
\end{align}

\subsection{Dynamic equations in complex form without nonlocal operators}
\label{sec:nonnonlocaloparator}

Both Eqs. \e{Reqn},\e{Veqn} and  \e{ReqnQ}-\e{VeqnQ} involves  $\hat P^-$ which is the nonlocal operator. Sometimes for analytical study and looking for the explicit solutions one may need to avoid such nonlocal operator. To do that we use Eq. \e{zKinematic} with r.h.s. rewritten through Eq. \e{PiP} which gives  %
\begin{equation} \label{zKinematiccomplex}
z_t\bar z_u-\bar z_tz_u=\bar \Pi_u-\Pi_u
\end{equation}
for the kinematic BC in the complex form.

To satisfy the dynamic BC we use Eq. \e{psiexpl2alpha}, where the term $\hat {\mathcal H}\left [\frac{1}{|z_u|^2}\hat {\mathcal H}\psi_u\right ]$ is expressed through the complex conjugate of Eq. \e{ztexplicit} and Eqs. \e{Projectordef} which results in%
\begin{equation} \label{HJ1}
\hat {\mathcal H}\left [\frac{1}{|z_u|^2}\hat {\mathcal
H}\psi_u\right ]=\frac{\bar z_t}{\bar z_u}-\I\frac{1}{|z_u|^2}\hat
{\mathcal H}\psi_u.
\end{equation}
Plugging in Eq. \e{HJ1} into Eq. \e{psiexpl2alpha}
and using Eqs. \e{Projectordef}
we obtain that

\begin{equation} \label{psiexpl2alpha2}
\psi_t=\psi_u \frac{\bar z_t}{\bar z_u}-\frac{1}{|z_u|^2}2\I\hat
P^-\left [\psi_u\hat {\mathcal H} \psi_u  \right
]-gy+\frac{\alpha(x_uy_{uu}-x_{uu}y_u) }{|z_u|^3}.
\end{equation}
We now note that using Eqs. \e{Hfpm} and \e{zPi}  allows to write
that   $\hat P^-\left [\psi_u\hat {\mathcal H} \psi_u  \right
]=\frac{\I}{4}\hat P^-\left [\bar \Pi_u^2-\Pi_u^2  \right
]=-\frac{\I}{4}\Pi_u^2$ thus reducing Eq. \e{psiexpl2alpha2} to
\begin{equation} \label{psiexpl2alpha3}
\psi_t\bar z_u-\psi_u \bar z_t+\frac{\Pi_u^2 }{2z_u}+\frac{g}{2\I}\bar z_u(z-\bar z)+\frac{\I\alpha\bar z_u }{2|z_u|}\left (\frac{z_{uu}}{z_u}-\frac{\bar z_{uu}}{\bar z_u} \right )=0,
\end{equation}
where we also expressed gravity and surface tension terms through
$z$ and $\bar z$ using Eqs.  \e{zPi}.    Eq. \e{psiexpl2alpha3} for
the particular case $g=\alpha=0$ was  first derived in Ref.
\cite{ZD2012} (except there are trivial misprints in Eq. 3.54 of
that Ref.). Eq. \e{psiexpl2alpha3} is the complex version of
Bernouilli equation. Using Eqs.  \e{zPi} one can also express $\psi$
in Eq. \e{psiexpl2alpha3} through $\Pi$ and $\bar \Pi$ which gives a
fully complex form of Bernouilli equation as follows

\begin{equation} \label{psiexpl2alpha3}
(\Pi_t+\bar \Pi_t)\bar z_u-(\Pi_u+\bar \Pi_u)\bar z_t+\frac{\Pi_u^2 }{z_u}-\I g \bar z_u(z-\bar z)+\frac{\I\alpha\bar z_u }{|z_u|}\left (\frac{z_{uu}}{z_u}-\frac{\bar z_{uu}}{\bar z_u} \right )=0.
\end{equation}
Eqs. \e{zKinematiccomplex} and \e{psiexpl2alpha3}
are the dynamic equations in the complex form. They are not resolved with respect to the time derivative but they do not contain any nonlocal operator.
\section{Generalized hydrodynamics and integrability}
\label{ref:Generalizedhydrodynamicsintegrability}

We notice that all expressions derived in Section \ref{sec:Hamiltonian formalizm}
starting from Eq. \e{symplecticlike1} and in Section \ref{DynamicalequationsComplex} before Eq. \e{HamiltzPi} are valid for arbitrary Hamiltonian $H.$ In this Section we go beyond the standard Hamiltonian \e{HamiltzPi} to apply our Hamiltonian formalism for other physical systems beyond the Euler equations with free surface, gravity and surface tension. We call the corresponding dynamical equations by ``generalized hydrodynamics''.

The new Hamiltonian is written as%
\begin{equation} \label{Hgener}
H=H_{Eul}+\tilde H,
\end{equation}
where $H_{Eul}$ is the standard Hamiltonian \e{HamiltzPi} and%
\begin{equation} \label{Htilde}
\tilde H=\frac{\I \beta}{8}\int (z_u+\bar z_u-2)(z-\bar z)\D
u=\frac{\beta}{2}\int y\hat {\mathcal H} y_u \D u
\end{equation}
is the ``generalized" part which adds up to the potential energy. Here $\beta$ is the real constant. Using FT \e{FTdef}, one can also rewrite Eq. \e{Htilde} through Parseval's identity as%
\begin{equation} \label{Htildek}
\tilde H=-\frac{\beta}{2}\int|k||y_k|^2dk
\end{equation}
which shows that $\tilde  H$ is the sign-definite quantity. Here we
also used that the Hilbert operator $\hat {\mathcal H}$ turns into a
multiplication operator under FT as  $ (\hat {\mathcal H}_u
f)_k=\I\, \text{sign}{\,(k)}\,f_k$ which follows from Eqs.
\e{Projectordef} and Appendix \ref{sec:AppA}. Thus the additional
potential energy $\tilde H$ is positive for $\beta<0$ and negative
for $\beta>0.$

There are several physical interpretation of $\tilde H$. First case
$\beta>0$ corresponds e.g. to the dielectric fluid with a charged
and ideally conducting free surface in the vertical electric field
\citep{Zubarev_JETPLett_2000,Zubarev_JETP_2002,ZubarevJETP2008eng}.
Such situation is realized on the charged free surface of
a superfluid Helium \citep{ColeCohenPRL1969,ShikinJETP1970}.   Then
Eq. \e{Htilde} is valid   provided surface charges fully screen the
electric field above the fluid free surface. This limit was first realized
experimentally  in Ref. \citet{EdelmanUFN1980}.  Negative sign of
$\tilde H$ implies instability due to the presence of the electric
field. Another application occurs for the quantum Kelvin-Helmholtz
instability of counterflow of two components of superfluid Helium
\citep{LushnikovZubarevPRL2018}.   Second case $\beta<0$ corresponds
e.g. to the dielectric fluid with a free surface in the horizontal
electric field
\citep{ZubarevZubarevaTechPhysLett2006,ZubarevZubarevaTechPhysLett2008,ZubarevKochurinJETPLett2014eng}
and references therein. Positive sign of $\tilde H$ implies a
stabilizing effect of the horizontal electric field. Similar effects
can occur in magnetic fluids.  See
\citet{ZubarevJETP2008eng,ZubarevKochurinJETPLett2014eng,LushnikovZubarevPRL2018}
for more references on physical realizations of the generalized
hydrodynamics.

We now consider the dynamics Eqs. \e{Reqn0}, \e{Veqn0} for the Hamiltonian \e{Hgener},\e{Htilde}. Then $U$ is still given by Eq. \e{Udef2} according to Eq. \e{Udef1} because $\tilde H$ does not depend on $\Pi$. Eq. \e{Pintdef} results in
\begin{equation} \label{Pintdef2g}
\mathcal{P}=-\I g(z-w)-2\I\alpha  \hat P^-(Q_u\bar Q-Q\bar Q_u )+\beta\hat P^{-}(R\bar R-1),
\end{equation}
while % $U$ and
$B$ remain the same as in Eqs. \e{Udef2} and \e{Bintdef2} because the definitions \e{Udef1} and \e{Bintdef} involve only variations over $\Pi$ and $\bar\Pi$.

Eqs.   \e{Reqn0},\e{Veqn0},\e{Udef2},\e{Bintdef2},\e{Qdef} and  \e{Pintdef2g}
result in the generalization of Dyachenko Eqs. \e{Reqn},\e{Veqn} as follows

 \begin{align}
\frac{\partial R}{\partial t} &= \I \left(U R_u - R U_u \right), \label{Reqng}\\
\frac{\partial V}{\partial t} &= \I \left[ U V_u - R B_u -\beta R \hat P^{-}\frac{\p }{\p u}(R\bar R)\right ]+ g(R-1)-2\alpha R  \hat P^-\frac{\p }{\p u}(Q_u\bar Q-Q\bar Q_u ). \label{Veqng}
\end{align}

As a particular example until the end of this section we consider Eqs. \e{Reqng} and \e{Veqng} for $g=\alpha=0.$ We define $r$ as %
\begin{equation} \label{rdef}
r=R-1
\end{equation}
and linearizes Eqs.
\e{Udef2},\e{Bintdef2},\e{Reqng} and \e{Veqng} over small amplitude solutions in $r$ and $V$ which gives%
\begin{equation}\label{rVlin1}
\begin{split}
& r_t=-\I V_u, \\
& V_t=-\I\beta r_u,
\end{split}
\end{equation}
where we used that $r$ does not have zeroth Fourier harmonics implying $\hat P^-r=r$ and $\hat P^-\bar r=0.$ Excluding $V$ from Eq. \e{rVlin1} results in %
\begin{equation} \label{ruueq}
r_{tt}=-\beta r_{uu}.
\end{equation}
If $\beta=-s^2<0,$ $s>0$, then Eq. \e{ruueq} turns into a wave equation, %
\begin{equation} \label{rwave}
r_{tt}=s^2r_{uu},
\end{equation}
while for $\beta=s^2>0$ we obtain an elliptic equation.

We now go beyond a linearization and  consider fully nonlinear Eqs. \e{Udef2},\e{Bintdef2},\e{Reqng} and \e{Veqng} for $\beta=-s^2.$ We assume a reduction %
\begin{equation} \label{rVreduction1}
V=\I s r.
\end{equation}
Then Eqs. \e{Udef2} and \e{Bintdef2} result in $B=s^2\hat P^-(|r|^2) $ and $U=\I sr$. Plugging in these expressions into Eqs. \e{Reqng} and \e{Veqng}
results in a single equation%
\begin{equation} \label{r_tsingle}
r_t=sr_u,
\end{equation}
with a general solution %
\begin{equation}\label{rvs1}
\begin{split}
& r=f(u+st), \\
& v=\I sf(u+st)
\end{split}
\end{equation}
for the arbitrary function $f(u)$. This is a remarkable result because it is valid for arbitrary level of nonlinearity.
In a similar way, a reduction %
\begin{equation} \label{rVreduction2}
V=-\I s r
\end{equation}
in Eqs. \e{Udef2},\e{Bintdef2},\e{Reqng} and \e{Veqng} results in a single equation%
\begin{equation} \label{r_tsingle}
r_t=-sr_u,
\end{equation}
with a general solution %
\begin{equation}\label{rvs2}
\begin{split}
& r=g(u-st), \\
& v=-\I sg(u-st)
\end{split}
\end{equation}
for the arbitrary function $g(u).$

The existence of the general solutions  \e{rvs1} and \e{rvs2} for the reductions \e{rVreduction1} and \e{rVreduction2}, however, does not imply that one can obtain the explicit solution of the general Eqs. \e{Reqng} and \e{Veqng}
because a linear superposition of solutions \e{rvs1} and \e{rvs2}
is not generally a solution of Eqs. \e{Reqng} and \e{Veqng}.

We now consider the second case  $\beta=s^2>0$ and look at a reduction  %
\begin{equation} \label{rVreduction3}
V= s r.
\end{equation}
Then Eqs. \e{Udef2} and \e{Bintdef2} result in $B=s^2\hat P^-(|r|^2) $ and $U=s[r+2\hat P^-(|r|^2)$]. Plugging in these expressions into Eqs. \e{Reqng} and \e{Veqng}
results in a single equation (both equations for $r_t$ and $V_t$ coincide)%
\begin{equation} \label{r_tsingle3}
r_t=\I s\left( r_u [-1+2\hat P^-(|r|^2)]-(1+r)2\hat P^-(|r|^2)_u\right ).
\end{equation}
In a similar way, a reduction %
\begin{equation} \label{rVreduction4}
V= -s r
\end{equation}
in Eqs. \e{Udef2},\e{Bintdef2},\e{Reqng} and \e{Veqng} results in a single equation%
\begin{equation} \label{r_tsingle4}
r_t=-\I s\left( r_u [-1+2\hat P^-(|r|^2)]-(1+r)2\hat P^-(|r|^2)_u\right ).
\end{equation}
Eqs. \e{r_tsingle3} and   \e{r_tsingle4} interchange under a change of the sign of the time so it is sufficient to study one of them.

Infinite number of explicit solutions of Eqs. \e{r_tsingle3} and   \e{r_tsingle4} can be constructed.  We however do that indirectly by first considering the reduction \e{rVreduction3} for variables $z$ and $\Pi$ instead of $R$ and $V$.  We use Eq. \e{Zt} and its complex conjugate $\bar z_t=-\I \bar U\bar z_u$ together with Eqs. \e{Udef2} and \e{RVvar1} to obtain that %
\begin{equation} \label{ztzubar}
\I\left ( \frac{\bar z_t}{\bar z_u}-\frac{ z_t}{ z_u} \right )
=R\bar V+\bar R V=\frac{\bar V}{z_u}+\frac{V}{\bar z_u}.
\end{equation}
Eq. \e{rVreduction3} and its complex conjugate imply that %
\begin{equation}\label{VVbars}
\begin{split}
& V=s(R-1)=s\frac{1-z_u}{z_u}, \\
& \bar V=s(\bar R-1)=s\frac{1-\bar z_u}{\bar z_u}
\end{split}
\end{equation}
which allows to exclude $V$ and $\bar V$ from Eq. \e{ztzubar} resulting in the closed equation for $z$ as %
\begin{equation} \label{LGE0}
\I(\bar z_t z_u-z_t\bar z_u)=s(2-z_u-\bar z_u).
\end{equation}
A change of variables $z=G-\I s t$ in Eq. \e{LGE0} results in the
Laplace growth equation (LGE) given by \citep{Zubarev_JETPLett_2000,
Zubarev_JETP_2002,ZubarevJETP2008eng}
\begin{equation}\label{lge1} \mbox{Im}\,\left(\bar{G}_t
G_u\right)=-s.
\end{equation}
 LGE
is integrable in a sense of the existence of  infinite number of
integrals of motion
and its relation to the dispersionless limit of the
integrable Toda hierarchy
\citep{MineevWeinsteinWiegmannZabrodinPRL2000}.

One can also mention that LGE
was derived as the approximation  of
Hele-Shaw
flow (the ideal fluid pushed through a viscous fluid
in a narrow gap between two parallel plates), see Refs. \cite{Kochina_1945,GalinDoklaAkadNauk1945,ShraimanBensimonPRA1984,HowisonSIAMJApplMath1985,BensimonKadanoffLiangShraimanTangRevModPhys1986,MineevDawsonPRE1994}.
Also Ref.  \cite{CrowdyJFM2000} found that exact solutions for free-surface Euler flows
with surface tension (such as
Crapper's classic capillary water wave solutions \cite{CrapperJFM1957} and solutions of Refs. \cite{TanveerProcRoySoc1996,CrowdyStudApplMath2000,CrowdyPhysFluid1999}) are related to  steady solutions of
Hele-Shaw flows (with non-zero surface tension).

The reduction \e{rVreduction4} also results in LGE by the trivial change of sign in Eq. \e{lge1}. Similar to the case  $\beta=-s^2<0$ above, the existence of infinite number of solutions  for the reductions \e{rVreduction3} and \e{rVreduction4} in the case
 $\beta=s^2>0$ does not imply that one can obtain the explicit solution of the general Eqs. \e{Reqng} and \e{Veqng}
because a linear superposition of solutions of the corresponding LGEs
is not generally a solution of Eqs. \e{Reqng} and \e{Veqng}.
Nevertheless, we make a conjecture that the full system Eqs. \e{Udef2},\e{Bintdef2},\e{Reqng} and \e{Veqng} is integrable both for $\beta<0$ and  $\beta>0$.

\section{Conclusion and Discussion}
\label{sec:conclusion}

We derived the non-canonical Hamiltonian system \e{implectic2} which
is equivalent to the Euler equation with a free surface for general
multi-valued parameterization of surface by the conformal
transformation \e{zwdef}.   This generalizes the canonical
Hamiltonian system \e{Hamiltpsieta} of   Ref. \cite{Zakharov1968}
which is valid only for single-valued surface parameterization. The
Hamiltonian coincide with the total energy (kinetic plus potential
energy) of the ideal fluid in the gravitational field with the
surface tension. A non-canonical Hamiltonian system \e{implectic2}
can be written in terms of Poisson mechanics
\e{hamiltoniancanonicalpoisson}  with the non-degenerate Poisson
bracket \e{PoissonBracketsRDef}, i.e. it does not have any Casimir
invariant. That bracket is identically zero between any two
functionals of the canonical transformation \e{zwdef}. In future
work we plan to focus on finding of  integrals of motion which are
functional of that conformal map only so they will commute with each
other which might be a sign of the complete integrability of the
Hamiltonian system  \e{implectic2}. It was conjectured in Ref.
\cite{DZ1994}  that the system \e{Hamiltpsieta} is completely
integrable at least for the case of the zero surface tension. Since
then the arguments {\it pro} and {\it contra} were presented, see
e.g. Ref. \cite{DyachenkoKachulinZakharovJETP2013eng}. Thus this
question of possible integrability is still open and very important.

We also reformulated the  Hamiltonian system \e{implectic2} in the complex form which is convenient to analyze the dynamics in terms of analytical continuation of solutions into the upper complex half-plane. A full knowledge of such singularities would provide a complete description of the free surface hydrodynamics and corresponding Riemann surfaces as was e.g. demonstrated on the particular example of Stokes wave in Ref. \citet{ LushnikovStokesParIIJFM2016}.

Additionally, we  analyzed the generalized hydrodynamics with multiple applications ranging from dielectric fluid with free surface in the electric field to the two fluid hydrodynamics of superfluid Helium. In that case we identified powerful reductions which allowed to find general classes of particular solutions. We conjecture that the generalized hydrodynamics might be completely integrable.

Extension of $2D$ results of this paper into $3D$ is beyond the scope of this work. We only note that the Hamiltonian Eqs. \e{Hamiltpsieta}
for single-valued parameterization are valid in $3D$ also \cite{Zakharov1968}. Also multi-valued parametrization can be extended into $3D$ provided the variation of waves is slow in the third dimension as shown in Ref. \cite {RubanPRE2005quasiplanar}.

\section{Acknowledgements.}
 The  work of A.D., P.L.   and V.Z.  was   supported by the state assignment "``Dynamics of the complex materials"".
 The  work of
P.L.  was   supported by the National Science Foundation, grant
DMS-1814619. The work of V.Z. was supported by the National Science
Foundation, grant number DMS-1715323.
%The work of P.L. on conformal mapping was supported by
%the Russian Science Foundation, Grant No. 14-22-00259.

\appendix
%\section{}\label{appA}

\section{Projectors to functions analytic in upper and lower complex half-planes}
\label{sec:AppA}%\label{sec:Projectors}

% ???(one can also assume that $q_u(u)$ % is piecewise continuous with at most a finite number of jumps ??? to ensure %that Eq. \e{fpm0} is valid, i.e. inverse FT recovers $q(u)$ pointwise).  Analytical continuation of $q^+(u)$ and $q^-(u)$ into complex $w$ straightforwardly  reduces to replacing $u$ by $w$ in Eqs. \e{fplus} an \e{fminus}. Then $q^+(w)$ and $q^-(w)$ are holomorphic in    $\mathbb{C}^+$ and  $\mathbb{C}^-$, respectively, by the converge of integrals \e{fplus} an \e{fminus}.

This appendix justifies the definitions    \e{Projectordef} of  the projector operators $\hat P^{\pm}$ as well as provides a derivation of Eqs. \e{xHy} and \e{PhiTheta1}. The Sokhotskii-Plemelj theorem  (see e.g.
\cite{Gakhov1966,PolyaninManzhirov2008}) results in
\begin{align} \label{sokhotskiip}
& \int\limits^{\infty}_{-\infty}\frac{q(u')du'}{u'-u+\I 0}=\text{p.v.}\int\limits^{\infty}_{-\infty}\frac{q(u')du'}{u'-u}-\I \pi  q(u)=\pi\hat {\mathcal H}q-\I \pi  q(u), \\
& \int\limits^{\infty}_{-\infty}\frac{q(u')du'}{u'-u-\I
0}=\text{p.v.}\int\limits^{\infty}_{-\infty}\frac{q(u')du'}{u'-u}+\I
\pi  q(u)=\pi\hat {\mathcal H}q+\I \pi  q(u), \label{sokhotskiim}
\end{align}
where we used the definition
\e{Hilbertdef} and $\I 0$ means $\I \epsilon, \ \epsilon\to 0^+$. Here  $q(u)\in \C$, $q(u)\to0$ for $u\to\pm \infty,$ as well as we assumed that $q(u)$ is H\"older continuous function, i.e.
$|q(u)-q(u')|\le C |u-u'|^\gamma$ for any real $u,$  $u'$ and
 constants $C>0,$  $0< \gamma\le1.$  The nonzero limit   $q(u)\to q_0=const\in \C$  at     $u\to\pm \infty$ also allows the convergence of integrals but of the decaying boundary conditions \e{xyinfinity} and \e{Dirichlet1} ensures that $q_0=0.$ To ensure a finite value   of $\hat {\mathcal H} q$ in Eq. \e{sokhotskiim} we also assume that a decay condition%
\begin{equation} \label{infdecay}
|q(u)-q_0|\le A|u|^{-\gamma_1}
\end{equation}
for $u\to\pm \infty$  with the constant values $\gamma_1>0, \ A>0$.
 The H\"older
continuity requirement  is not necessary for applicability of Eqs.
\e{sokhotskiip} and \e{sokhotskiim} and can be relaxed  (see e.g.
\cite{TitchmarshBook1948,Gakhov1966,PandeyBook1996}). E.g., instead of the H\"older
continuity one can assume that  $q\in L^p$  then $\hat {\mathcal H}
q\in L^p$ for any $p\in (1,\infty)$ with $\| q\|_{L^p}\equiv\left(
\int^\infty_{-\infty}|q(u)|^p\D u\right )^{1/p}.$ The condition
$q\in L^p$ is sufficient for the existence of the inverse of $\hat
{\mathcal H}$ such that $\hat {\mathcal H}^2 q=-q$ almost
everywhere.  The Hilbert transform can be also considered for
bounded almost everywhere functions $q\in L^\infty$ which implies
that  $\hat {\mathcal H} q$ belongs to the  bounded mean oscillation
(BMO) classes of functions
\citep{FeffermanBullAmerMathSoc1971,FeffermanSteinActaMath1972}.
However, H\"older continuity requirement and the decay condition
\e{infdecay} are typically sufficient for our purposes as well as
they ensures that $\hat {\mathcal H}^2 q=-q$ pointwise. E.g. a
singularity of a limiting Stokes wave $\propto u^{2/3}$
~\citep{Stokes1880} corresponds to $\gamma=2/3$. The limiting
standing wave is expected to have a singularity  with $\gamma=1/2$
\citep{PenneyPricePhilTransRSocA1952,MalcolmGrantJFM1973StandingStokes,WilkeningPRL2011}.
Generally in this paper, $q(u)$    is formed from functions analytic
at the real line $w=u$ and their complex conjugates. It implies that
typically  $\gamma=1.$ Only in exceptional cases,  complex
singularities reach $w=u$      from $w\in \C$ implying that
$\gamma<1$ as for the limiting Stokes wave and limiting standing
wave.

Using Eqs. \e{sokhotskiip} and \e{sokhotskiim}, we rewrite Eq. \e{Projectordef} as follows
\begin{align} \label{Puminus}
\hat P^{\pm}q=\frac{1}{2}(1\mp\I \hat {\mathcal
H})q=\pm\frac{1}{2\pi
\I}\text{p.v.}\int\limits^{\infty}_{-\infty}\frac{q(u')du'}{u'-u}+\frac{1}{2}q(u)=\pm\frac{1}{2\pi
\I}\int\limits^{\infty}_{-\infty}\frac{q(u')du'}{u'-u\mp\I 0}.
\end{align}Extending $u$ into
the complex plane of $w$ in Eqs. \e{Puminus} either in $\C^+$  or in
$\C^-$  (one can also interpret that as closing complex integration
contours  in $\C^+$  or in  $\C^-$) we obtain that
\begin{equation} \label{pplusdef}
q^+\equiv \hat P^+q
\end{equation}
is analytic in  $\C^+$  and
\begin{equation} \label{pminusdef}
q^-\equiv \hat P^{-}q
\end{equation}
is analytic in  $\C^-$ such that $q^\pm(u)\to 0$ for $u\to\pm \infty$. Using Eqs. \e{Puminus}-\e{pminusdef}
we obtain that
\begin{equation} \label{qprojectiondefAppendix}
q=q^++q^-.
\end{equation}
Eqs. \e{pplusdef}-\e{qprojectiondefAppendix}  justify the definition    \e{Projectordef} of  $\hat P^{\pm}$
as the projector operators as well as Eq. \e{qprojectiondef} if we keep in mind that $q_0=0$ for all functions of interest because of the decaying boundary conditions \e{xyinfinity} and \e{Dirichlet1}.
We note that  Eqs. \e{Pfm} can be also immediately obtained by plugging Eq. \e{qprojectiondefAppendix} into Eqs. \e{Puminus} and moving integration contour from the real line $u=w$   either upwards into   $\C^+$  or downwards into    $\C^-.$

Assume that $q(w)$ is the analytic function for $w\in\C^-,$ i.e. $q^-\equiv 0$ in Eq. \e{qprojectiondefAppendix}.   Moving the integration contour in Eq. \e{sokhotskiim}  from the real line $u=w$   downwards into    $\C^-$   implies  the zero value of the integral. Then taking the real and imaginary parts of r.h.s. of Eq. \e{sokhotskiim}, i.e. setting $\hat {\mathcal H}q+\I \pi  q(u)=0$,   results in the relations between real and imaginary parts of $q$ at the real line $w=u$ as follows (\cite{Hilbert1905})%
\begin{equation} \label{ReImHilbert}
\hat {\mathcal H} Re(q)=Im(q), \quad \hat {\mathcal H} Im(q)=-Re(q).
\end{equation}
We also notice that Eqs. \e{xHy} and \e{PhiTheta1} are  obtained from Eqs. \e{ReImHilbert}   if we set either $q(w,t)=z(w,t)-w$ or $q=\Pi(w,t)$ which ensures that $q(w,t)$ is analytic for $w\in \C^-$.

 Another view of the projector operators $\hat P^{\pm}$ can be obtained if we use the Fourier transform (FT)  %
\begin{equation} \label{FTdef}
q_k\equiv \frac{1}{(2\pi)^{1/2}}\int \limits
^{\infty}_{-\infty}q(u)\exp\left (-\I ku\right )\D u
\end{equation}and introduce the splitting of   $q(u)$ as
\begin{equation} \label{fpm0}
q(u)=q^+(u)+q^-(u),
\end{equation}
where
\begin{equation} \label{fplus}
q^+(w)=\frac{1}{(2\pi)^{1/2}}\int \limits ^{\infty}_{0}q_k\exp\left
(\I kw\right )\D k
\end{equation}
is the analytical (holomorphic) function  in $\mathbb{C}^+$ and
\begin{equation} \label{fminus}
q^-(w)=\frac{1}{(2\pi)^{1/2}}\int \limits ^{0}_{-\infty}q_k\exp\left
(\I kw\right )\D k
\end{equation}
is the analytical function  in $\mathbb{C}^-$. Here we assume that
the  inverse FT, $$\mathcal
F^{-1}[q_k](u)\equiv\frac{1}{(2\pi)^{1/2}}\int \limits
^{\infty}_{-\infty}q_k\exp\left (\I
ku\right )\D k,$$ equals almost everywhere to $q(u)$ for  real values of $u. $ This is valid e.g. if  $q(u)$ belongs to both  $ L^1$ (absolutely integrable)  and $ L^2$ (square integrable) classes (see e.g. Ref. \citet{RudinBookRealandComplexAnalysis1986}).
If the function $q(w)$ is analytic in $\C^-$ then    $\bar{q}({w})$
 is analytic in $\C^+$ as also seen from equations \e{fpm0}-\e{fminus}.

%\bibliographystyle{jfm}
%\bibliography{surfacewaves,lushnikov,biblionls,books,helium}

\end{document}